\documentclass[11pt,a4paper]{article}
\usepackage{jinstpub}
\pdfoutput=1

\usepackage[colorinlistoftodos]{todonotes}
\usepackage[section]{placeins}
\usepackage[nottoc]{tocbibind}
\usepackage{caption}
\usepackage{subcaption}
\usepackage{lineno}
\title{\centering Construction of Precision sMDT Detector for the ATLAS Muon Spectrometer Upgrade}

\author[a]{D. Amidei}
\author[a, 1]{N. Anderson}
\author[a]{A. Chen}
\author[a]{E.~Carpenter}
\author[a]{L. Cooperrider}
\author[a]{T. Dai}
\author[a]{E. Diehl}
\author[a,1]{C. Ferretti}
\author[a]{Y. Guo}
\author[a, c] {J. Li}
\author[a]{X. Meng}
\author[a]{K. Nelson}
\author[a]{V. Pillsbury}
\author[a]{E. Salzer}
\author[a]{T. Schwarz}
\author[a]{L. Simpson}
\author[a]{Z. Wang}
\author[a]{C. Weaverdyck}
\author[a, b]{C. Wei}
\author[a, b]{Z. Yang}
\author[a]{M. Yuan}
\author[a]{B. Zhou}
\author[a]{J. Zhu}

\affiliation[a]{Department of Physics, 
University of Michigan, 450 Church Street, 
Ann Arbor, MI 48109,  USA}
\affiliation[b]{Department of Modern Physics, 
University of Science and Technology (USTC), 96, 
JinZhai Road Baohe District, Hefei, Anhui, 230026, 
China}
\affiliation[c]{Physics Department, Shanghai 
Jiao-Tong University, 800 Dongchuan Road, 
Minhang, Shanghai, China}
\emailAdd{claudio.ferretti@cern.ch, %,neal.anderson@cern.ch, 
Contact Editor}

\abstract{

This paper describes the small-diameter monitored 
drift-tube detector construction at the 
University of Michigan as a contribution to 
the ATLAS Muon Spectrometer upgrade for the 
high-luminosity Large Hadron Collider at CERN.
Measurements of the first 30 chambers built at
Michigan show that the drift tube wire position
accuracy meets the specification of 20~$\mu$m. 
The positions of the platforms for alignment and 
magnetic field sensors are all installed well within 
the required precision. 
The cosmic ray test measurements show single 
wire tracking resolution of 
$\rm 100 \pm 7 \; \mu m $
with an average detection efficiency
above $\rm 99$\%.
The infrastructure, tooling, techniques, and
procedures for chamber production are described
in detail. 
The results from the chamber quality control tests 
of the first 30 constructed chambers are reported.
}

\begin{document}
\maketitle
\flushbottom
%
%------------------------------
\section{Introduction}\label{sec:intro}
\begin{figure}[!ht]
    \centering
   \subfloat[]{
    \includegraphics[width=0.8\textwidth]{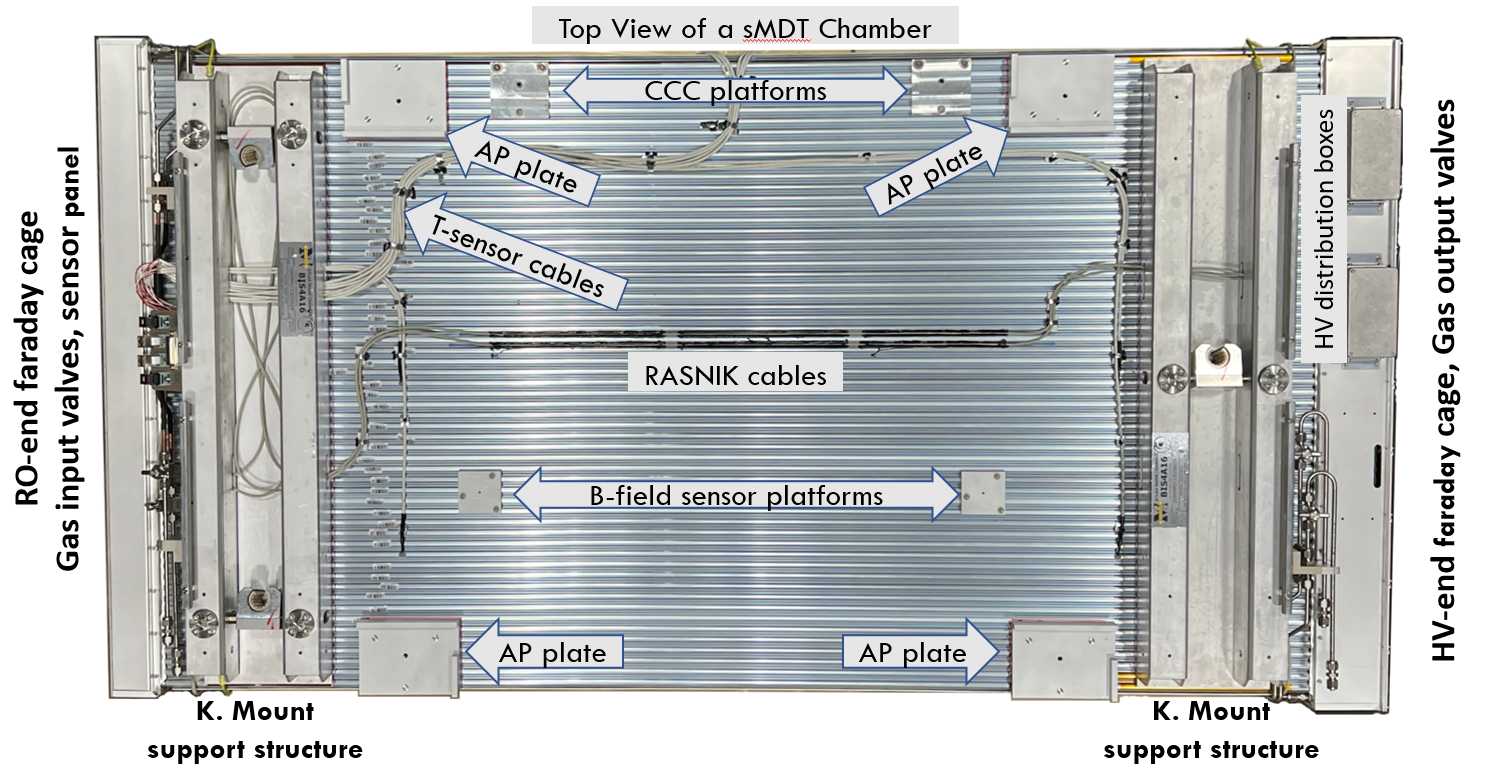}
    }
 \vspace{0.26cm}
    \subfloat[]{
    \includegraphics[width=0.75\textwidth]{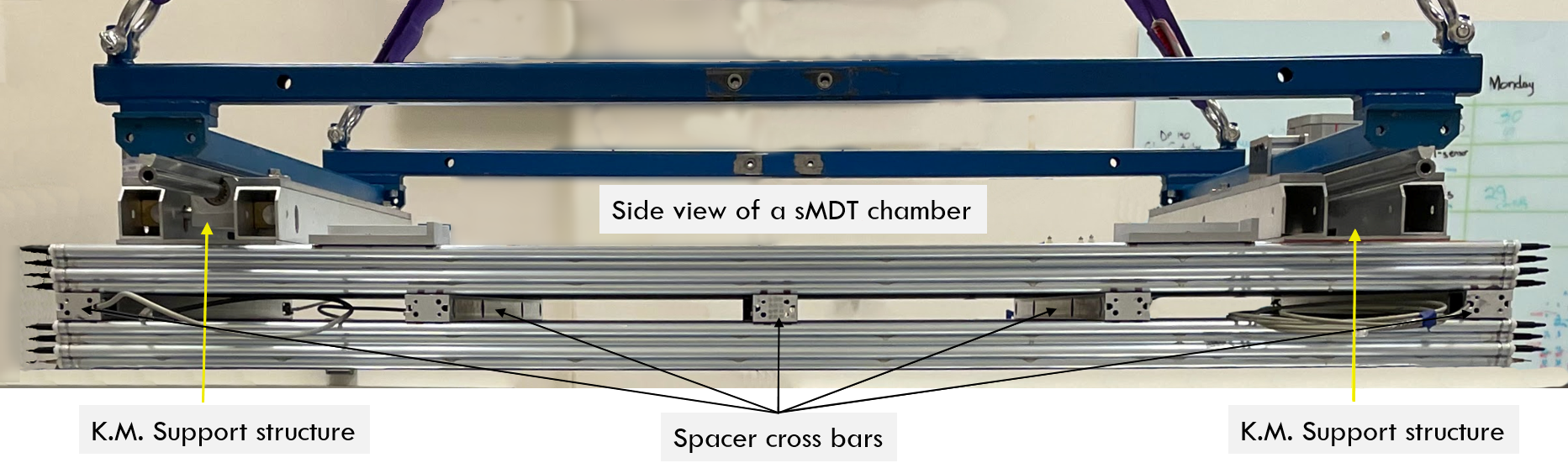}
    }
    \caption{Chamber view: a) {\sl top} with
    platforms, support structures, temperature
    cable routing, sensors readout panel, 
    gas pipes, Faraday cages, and HV splitter 
    boxes.
    CCC and AP platforms are intended for the 
    global alignment sensors;
    (b) {\sl side} with kinematic mounts 
    on the support structures holding the chamber
    to a overhead crane. The spacer bars are 
    visible between the two ML's.}
    \label{fig:sMDT-draw}
\end{figure}

%\Section{Introduction}
The ATLAS Muon Spectrometer (MS) is the largest 
muon spectrometer ever 
constructed~\cite{AtlasCollaboration_2008}.
Currently in the barrel 
(pseudo-rapidity $|\eta| < $ 1.05)
Resistive Plate Chambers (RPC) 
provide fast timing trigger-signals to 
identify events containing muons
and Monitored Drift-Tube (MDT) chambers
provide precision tracking.
In the end-cap 
(1.05 $< |\eta| < $ 2.7) the muon trigger
comes from Thin Gap Chambers (TGC) 
and precision tracking from MDTs and
in the innermost region the new 
Small-Wheel~\cite{NSWtdr} (consisting of 
MicroMegas and small-gap TGC detectors). 
The barrel contains 20 layers of MDTs in 
three concentric stations, each with
a single tube resolution of 80~$\mu$m, 
plus 3 stations of RPCs on the 
middle and outer MDT stations.  
The MDTs are arranged in 16 sectors 
around the beam line, covering $2\pi$
in alternating "large" and "small" chamber sizes,
referred to as "Large" and "Small" sectors.  
The ATLAS MS will be upgraded ~\cite{phase2TDRMuonSpectrometer}, 
scheduled in 2026-28,
for the high-luminosity LHC (HL-LHC) operations. 
One major MS upgrade project is to add
an extra triplet-station of small-gap RPC (sRPC) 
chambers to the barrel inner MS station to 
improve the muon Level-1 trigger efficiency.
To accommodate the new sRPCs, 
the MDTs of 8 Barrel-Inner-Small (BIS) 
sectors will be replaced with small-diameter 
(15 mm) MDTs (sMDT's)
designed and developed by the Max 
Planck Institute (MPI)~\cite{MDTforHLLHC},
which are half of the thickness of the MDTs.  
\begin{figure}[!ht]
    \centering
    \includegraphics[width=0.75\textwidth]{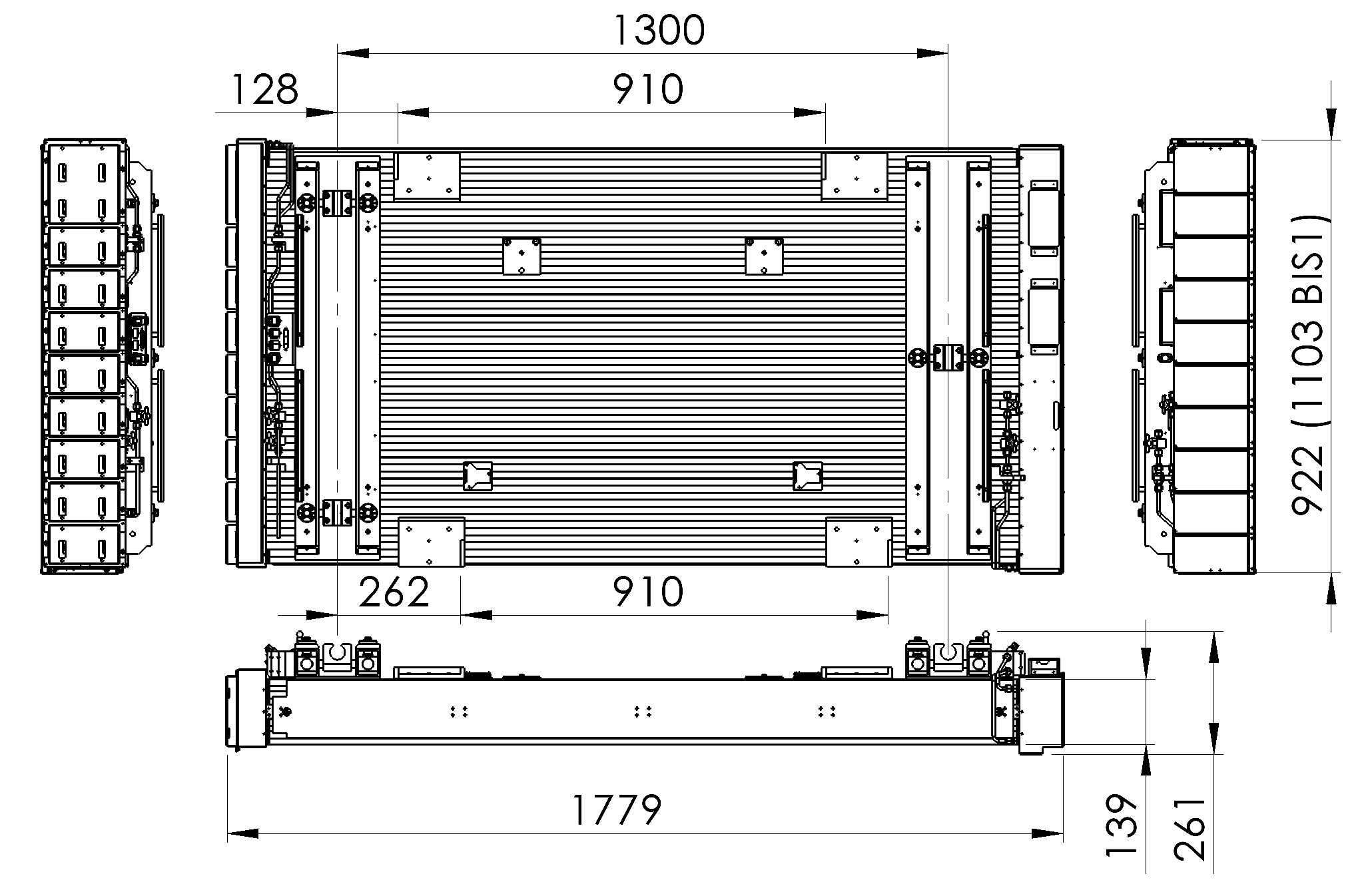}
\caption{Drawing of a BIS2 sMDT chamber.
Quoted dimensions are in mm.}
\label{sMDTdraw}
\end{figure}
The 8 Large sectors have space for inserting 
the new sRPCs, so the current MDTs will remain.
MPI and the University of Michigan (UM) 
are sharing equally the production 
of a total  of 100 sMDT chambers.
An sMDT chamber consists of 2 tube 
multi-layers (ML), each containing 4 tube 
layers of 70 (BIS1) or 58 (BIS2-6) tubes.
The geometry of a BIS2 is shown
in figure~\ref{sMDTdraw}, while
figure~\ref{fig:sMDT-draw} shows a complete 
BIS4 chamber built at UM. 

For the US site, the tubes are constructed 
and tested at Michigan State University (MSU), 
then shipped to UM, where they are 
fully retested~\cite{UMtubePaper} 
before chamber construction.
A spacer frame (parts made at IHEP\footnote{Institute 
for High Energy Physics, Protvino, Russia}), 
containing the RASNIK in-plane alignment 
system~\cite{Beker_2019}
(from NIHKEF\footnote{National Institute 
for Subatomic Physics,Amsterdam, Netherlands}), 
is glued between the two ML's. 
Platforms (from Saclay\footnote{Saclay Nuclear
Research Centre, Saclay, France}) for mounting 
the magnetic field sensors and global alignment 
devices~\cite{OpticATLASmuon}
are glued on top of the chamber, together with 
the chamber kinematic mounts on supporting
structures (from IHEP).
Temperature sensors assembled with cables 
(from NIHKEF) are distributed on both ML's.
The Faraday cages (from IHEP) cover the front-end 
electronics (from LMU\footnote{Ludwig Maximilian 
University, Munich, Germany}) with 2 HV 
distribution boxes (from MPI) mounted on top 
of them.

This paper reports the sMDT chamber construction 
infrastructure, assembly procedure, quality control, and performance test results at UM.

%------------------------------

\section{Infrastructure and Tooling for sMDT Construction and Testing}
\label{sec:infrastucture}
From 2017 to 2020 UM built the infrastructure for 
sMDT chamber construction as well as two prototype 
precision sMDT chambers to certify the tooling and 
chamber assembly procedures. 
The mass production of the sMDT chambers at UM
started in April 2021.
 The sMDT chamber construction 
infrastructure and tooling are described in this section.

\subsection{High bay and large granite table}

\begin{figure} [hbt]
    \centering
    \subfloat[]{
    \includegraphics[width=.42\textwidth]{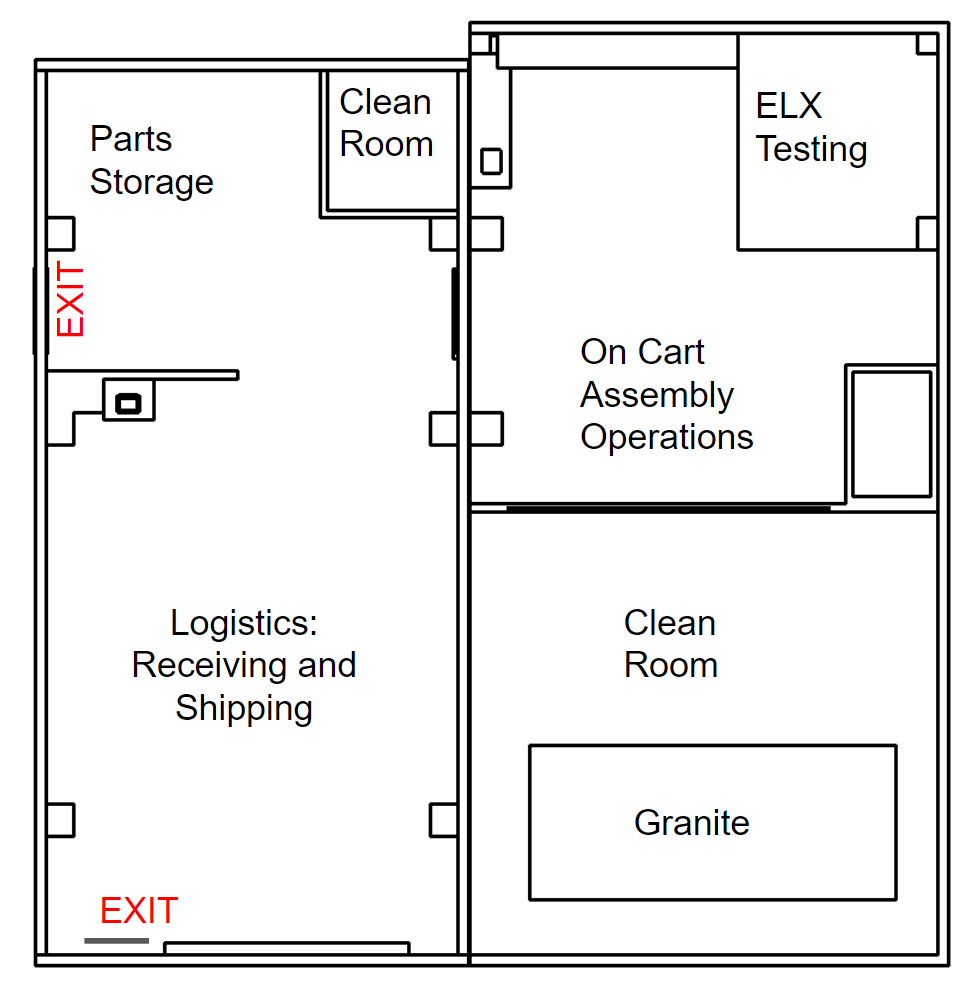}
    }
    \subfloat[]{
    \includegraphics[width=.56\textwidth]{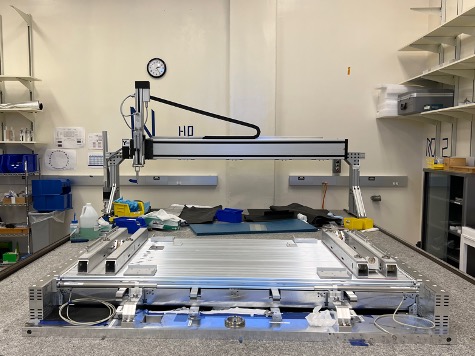}}
    \caption{(a) Map of the sMDT construction labs 
    in the high bay area.
    (b) The glue-machine and a constructed sMDT 
    chamber on the granite table in the 
    temperature and humidity controlled room.}
    %{\color{red}CHANGE CAPTION - DONE/check}
    \label{fig:my_label}
\end{figure}

The chamber construction makes use of 2 large 
($20\times20$~m$^2$) high bay research labs
(see figure~\ref{fig:my_label}(a))
in the UM physics building.  
These are equipped with multiple overhead cranes
and a clean room portion outfitted with a 
$ 3 \times 7 $~m$^2$ granite table with 
25~micron flatness.
This clean room utilizes a dedicated
HVAC system to maintain ${}\pm 0.5 ^\circ C $
temperature stability and ${}\pm 5 \%$ humidity 
stability.
The chamber parts (tubes, 
structural elements and platforms) are assembled 
on the granite table using precision jigging and  
glued together using an automated epoxy dispensing
machine.

Installation of the chamber's gas system and 
front-end electronics is done after the chamber
is moved to mobile stations in the space outside
the clean room.
\par
A sub-room in the corner (labeled as "ELX 
Testing" in the room map) was built to keep 
the humidity below 40\% for chamber cosmic 
ray testing.
Another bay-area is used for storing and shipping 
chambers as it has two overhead cranes and a roll 
up outside door for loading chambers into a 
shipping container.
A small clean room in this bay area is used for 
clean parts storage and gas-bar pre-assembly.

\subsection{Automatic glue machine}

\begin{figure}[hbt]
    \centering
    \subfloat[]{
    \includegraphics[width=.42\textwidth]{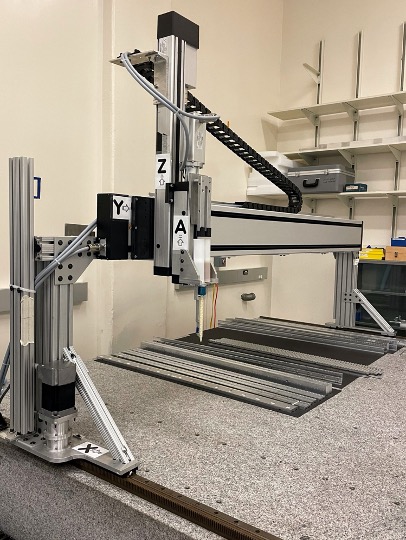}
    }
    \hspace{0.05cm}
    \subfloat[]{
    \includegraphics[width=.56\textwidth,angle=270]{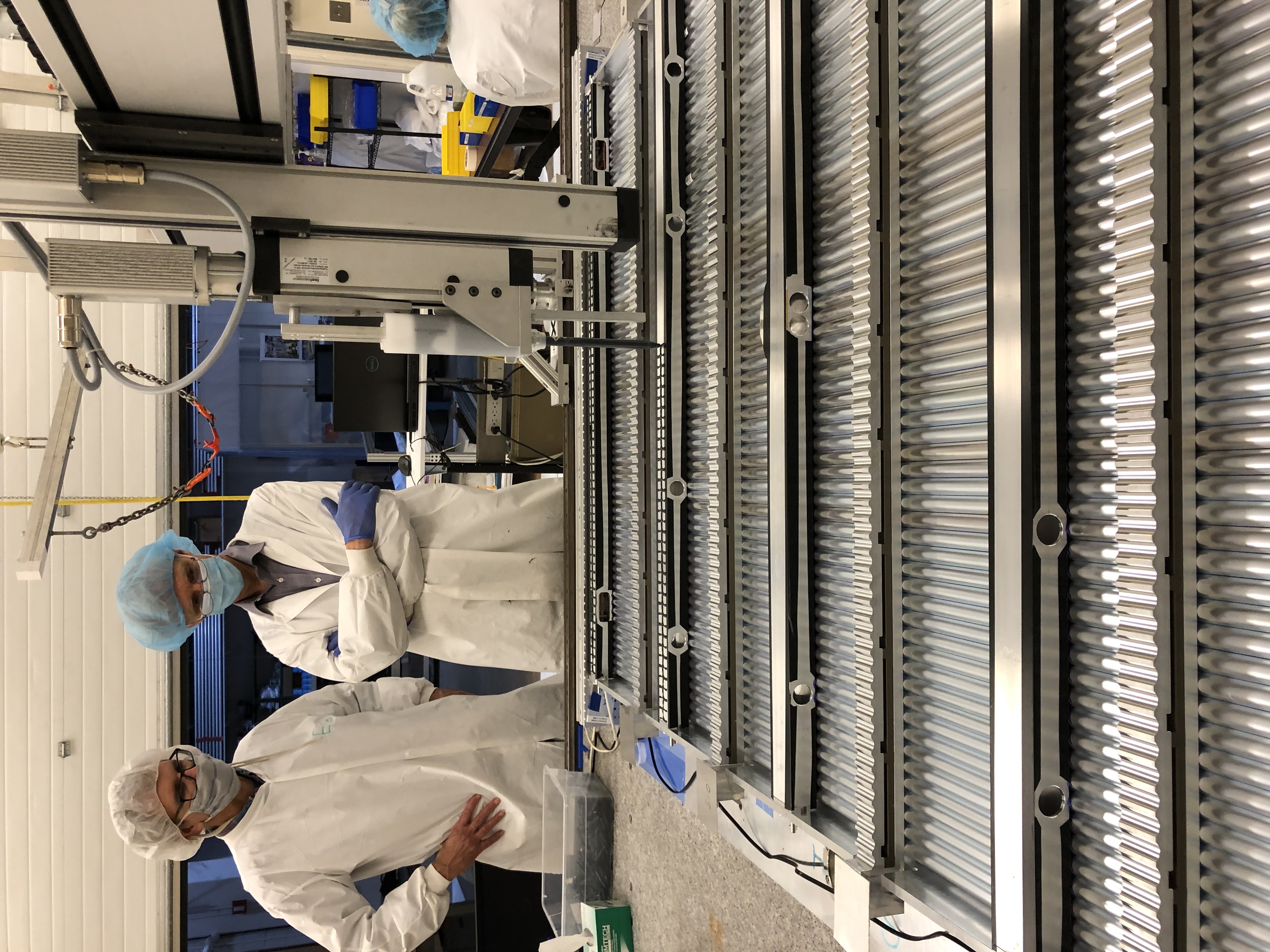}
    }
    \caption{
    (a) Four motion actuators for 3D motion of the 
    glue dispenser, all computer-controlled.
    (b) Gluing the spacer frame on the bottom multi-layer of a chamber.
    }
    \label{fig:Glue-machine}
\end{figure}

A large gantry-mounted, computer driven glue 
dispensing machine was designed and built at UM.
It controls the 3D motion (marked X, Y, and Z on 
the machine) of the dispensing tip as well as the 
drive piston for the epoxy cartridge (A drive).
The gantry is assembled from 
$\rm 7.6 cm  \times 15.2 cm $ 
aluminum extrusions which ride on 
carriages on rails mounted directly
on the surface of the granite table
(see figure~\ref{fig:Glue-machine}(a)).
The two rails on the granite are aligned to be parallel within 7 micro-radians. The X drive engages a toothed rail on one side of the granite, which moves the gantry 
along the long dimension of the table with a precision of 0.1~mm.
The linear rail carriages and rail are kept 
clean and lubricated to minimize frictional drag. 
The Y, Z, and A drives are linear 
actuators\footnote{https://www.isel.com/en/linear-units-les4.html, and https://www.isel.com/en/linear-units-les5.html}.
The gantry Y axis, with a range of 2.5~m,
travels along the length of the tubes 
(see figure~\ref{fig:Glue-machine}(b)). 
The actuator along the Z axis has a range of
0.5~m to allow the movement of the gluing 
cartridge over the whole vertical dimension 
of a chamber, from the first layer to the 
supporting structures at the top.
The A drive expels epoxy from the cartridge 
and is programmed to move in steps of 
0.26~mm to dispense 0.5~cc epoxy at a time.
The control software is written with LabView\cite{bitter2006labview} 
which interfaces with the motor drivers 
provided by the manufacturers.
Specific sequences of gantry motions and 
epoxy dispensed at each location 
on the gantry path are designed and tested 
to determine the configurations used for gluing 
each tube layer, spacer frame, and support 
structure.

\subsection{Precision combs}

\begin{figure}[!htb]
    \centering
    \subfloat[]{\includegraphics[width=.5\textwidth]{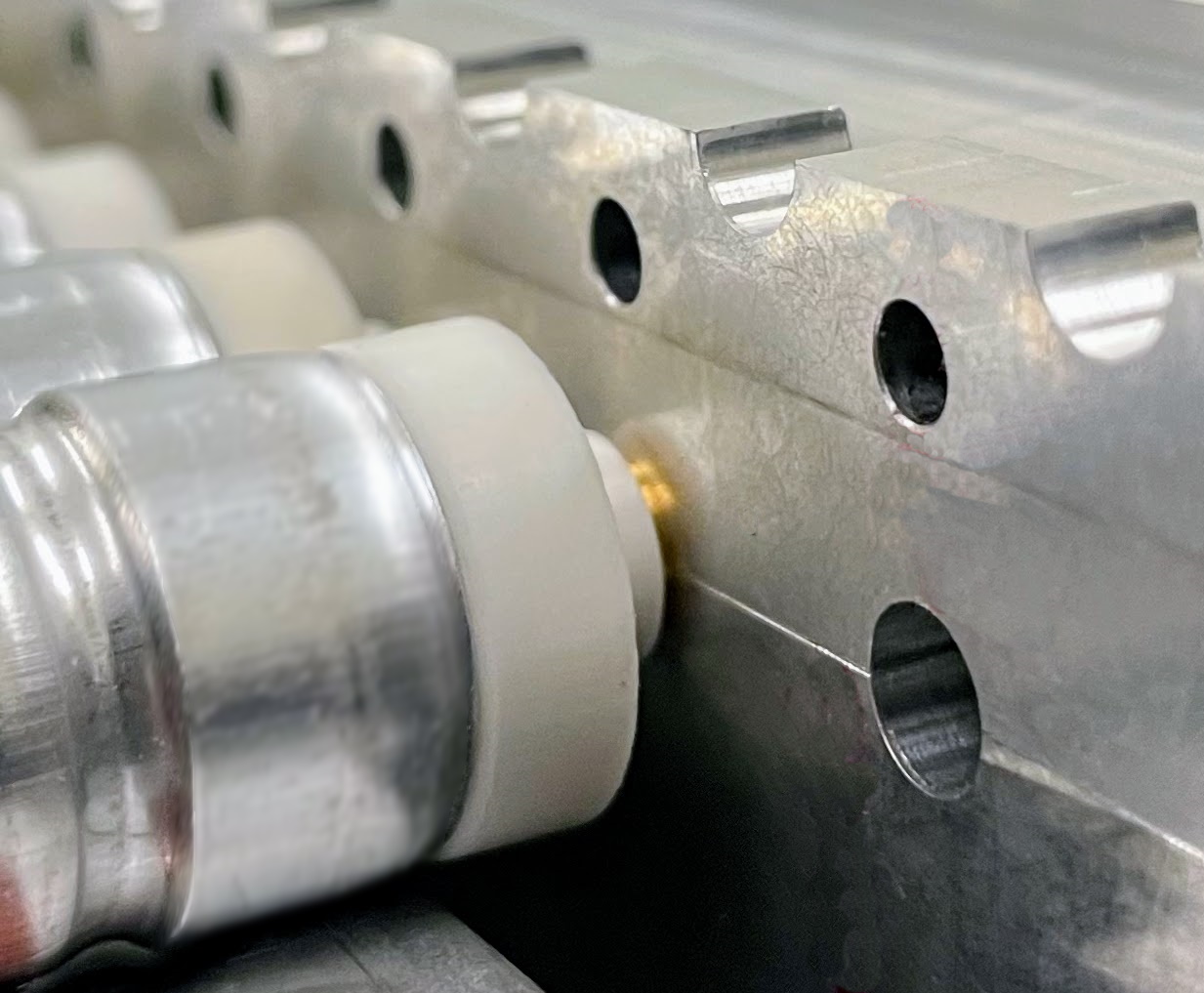}
    }
    %\hspace{0.1cm}
    \subfloat[]{\includegraphics[width=.48\textwidth]{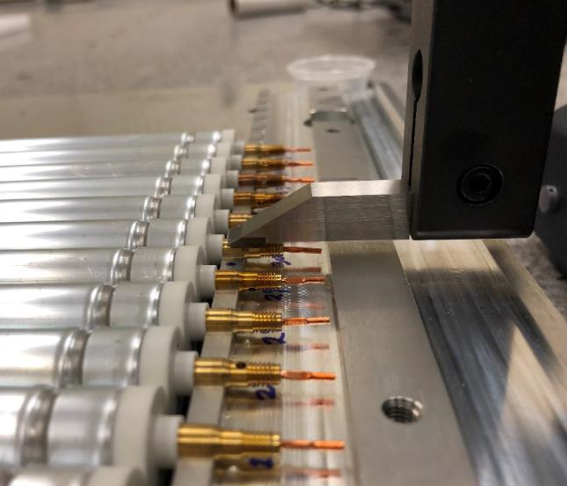}
    }
    \caption{
    (a) Upper and lower combs constrain the 
    location of the endplugs with high precision.
    The smaller holes a the top of the picture
    allow insertion of ground anchors which are screwed into the gaps between tubes after the epoxy has cured;
    (b) Tubes placed in the 
    chamber construction combs.
    The heights of the endplug precision cylinders are measured using a height gauge.
    }
    \label{fig:precision-jigging}
\end{figure}

The precision combs are the jigging used to 
precisely position tubes for chamber construction 
and were designed by MPI and made in Germany for 
both production sites.
Combs are used on both the chamber readout (RO) 
and high voltage (HV) ends.  Each endplug has 
a precision brass surface of diameter
$ 5.00 \, \genfrac{}{}{0pt}{1}{+0.00}{-0.01} $~mm 
which is concentric to the tube-center 
(therefore to the wire) within 5~microns rms. 
The endplug precision surface is captured between 
two half-holes in adjacent stacked combs. 
A full comb set consists of 9 stacked plates with 
precision holes (paired half-holes) of diameter 
$ 5.00 \, \genfrac{}{}{0pt}{1}{+0.01}{-0.00} $~mm 
on the interfaces between adjacent plates 
(see fig~\ref{fig:precision-jigging}(a)).
Each plate is fixed to the one underneath with 
17 screws tightened to 7.5~Nm torque. 
The base comb is 50 mm thick for rigidity and
subsequent combs are 13.077~mm thick except for
a 45.600~mm thick comb which sets the gap for 
the spacer frame.
\par
The combs locate the tubes on a 15.100~mm 
horizontal pitch, and a vertical pitch 
(tube layer) of 13.077~mm.
During assembly of the tubes into a chamber, the 
precision cylinder of each tube's endplugs are 
placed in the two end-combs as shown in
figure~\ref{fig:precision-jigging}(b). 
The lower comb's half-cylinders constrain the 
horizontal locations of the tubes' precision 
cylinders.
The upper comb is screwed to the lower comb to provide a vertical constraint on tube position.
%is provided by the upper comb 
%layer being screwed  tightly to the layer below.

Besides four base middle combs fixed to the 
table shown in figure~\ref{fig:base-combs}(b), 
there is another set of four temporary combs
used on top of the bottom ML when gluing 
the fifth tube layer on the spacer frame bars.
These temporary combs replace the four middle 
combs on the granite which are not accessible 
when gluing the top ML.
In addition there is a set of four weights with 
7.5~mm radius semi-circles cut in the bottom
surface which are placed on top of each freshly 
glued tube layer to control the straightness and 
potential bowing of the tubes.

\subsubsection{Precision jigging survey and setup on granite table}

\begin{figure}[htb]
    \centering
    \subfloat[]{
    \includegraphics[width=.60\textwidth]{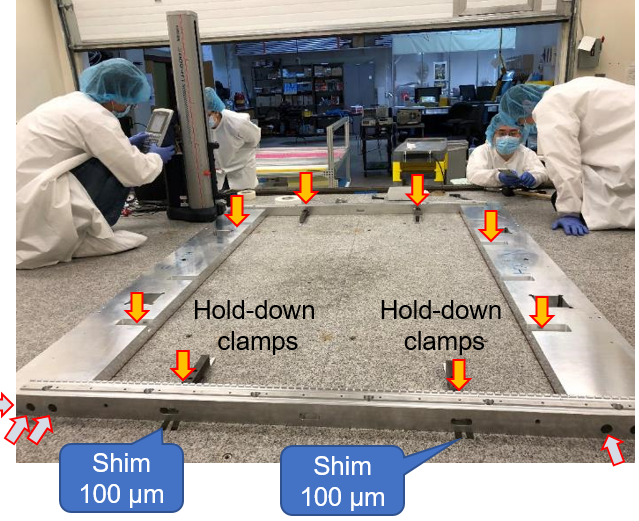}
    }
    \subfloat[]{
     \includegraphics[width=.37\textwidth]{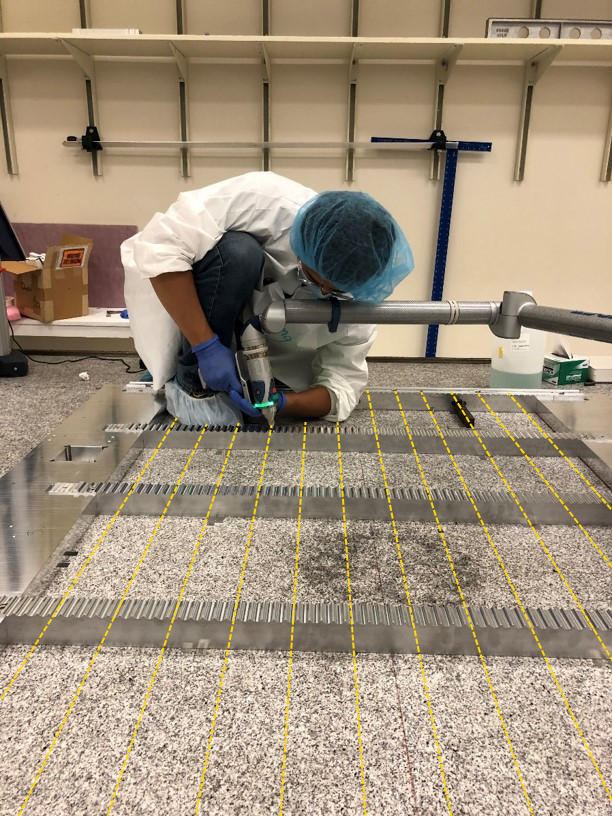}
     }
    \caption{Set up of base and mid-combs: 
    (a) the two precision base end-combs are    
        connected to two sidebars by bolts indicated by the white/red arrows.
        The yellow/red arrows point to the 
        hold-down clamps of the base jigging;
    (b) the middle base-combs are connected 
        to the two sidebars.
        }
    \label{fig:base-combs}
\end{figure}

The set up of the combs on the granite was
an iterative process, alternating between 
adjusting the jigging positions and making 
precision measurements of the jigging.
Positions were measured with both a 
FARO-arm\footnote{https://www.faro.com/en/Resource-Library/Tech-Sheet/techsheet-8-axis-edge-faroarm-scanarm} 
for 3D-measurements with precision of 10~$\mu$m
and a height gauge\footnote{https://www.mitutoyo.com/webfoo/wp-content/uploads/2104\_LH-600E.pdf}
for measuring height relative to the granite 
table surface with a precision of 5~$\mu$m.
Figure~\ref{fig:base-combs}(a) shows the two base 
precision end-combs which are bolted at opposite 
ends of two crossbars. The RO comb is in the 
foreground and the height gauge is visible
measuring the HV comb at the far side of the 
table.
The middle base-combs, shown in
figure~\ref{fig:base-combs}(b),
are connected to the two cross-bars and ensure tube straightness when tubes are glued.
The middle base-combs are aligned both along the tube direction (yellow-dashed lines in figure~\ref{fig:base-combs}(b)) and in height to better than 50~$\mu$m using the FARO-arm.
\par
Each end comb base is clamped onto the granite 
table at two points using 100~microns precision 
shims under the combs.  The combs are also
bolted to heavy side bars which set the length 
of the jigging.
The base combs need to be set parallel to each other 
and perpendicular to the wire direction. 
The wire direction must be aligned with the 
glue machine Y axis and this alignment is periodically verified and the Y gantry position tuned if necessary.

\begin{figure}[hbt]
    \centering
    \includegraphics[width=0.78\textwidth]{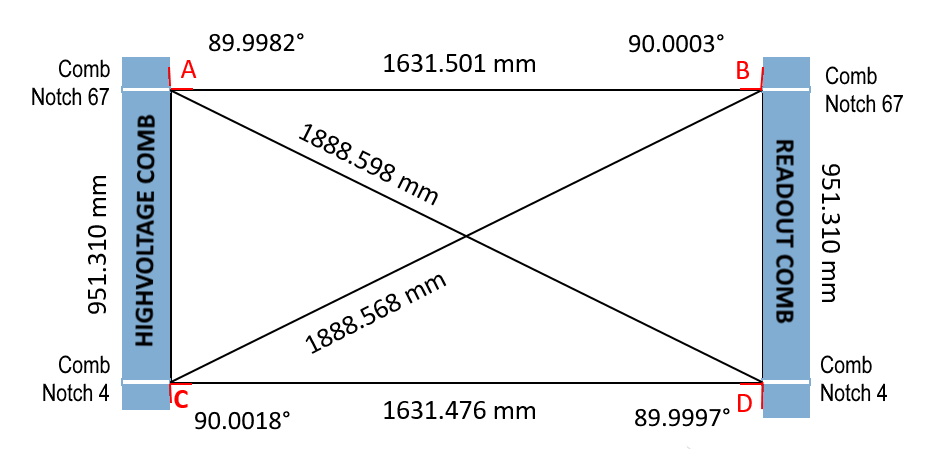}
    \caption{Base comb layout geometry and 
    measured distances and angles.}
    \label{fig:basediag}
\end{figure}

The squareness of the base-comb setup was checked
using the Faro-arm to measure the diagonals 
defined by precision half-circular cylinders 
placed on comb notches 4 and 67 (as in 
figure~\ref{fig:combgeometry-measurements}(a)). 
These diagonals differ by 15~microns from the 
average and the angles between the comb assembly
and the long base-plates have a maximum deviation
of 0.0018$^\circ$ = 0.03~mrad from 90$^\circ$,
as shown in figure~\ref{fig:basediag}.

\subsubsection{Measured precision of the combs}
\label{sec:measure-combs}

\begin{figure}[hbt]
    \centering
    \includegraphics[width=0.78\textwidth]{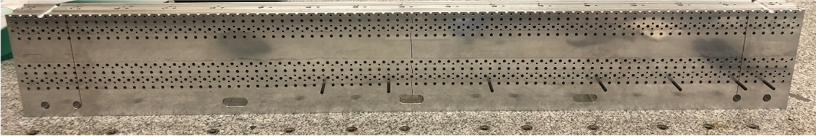}
    \caption{One set of the precision combs 
    stacked on the granite table.}
    \label{fig:oneset-comb}
\end{figure}

As mentioned previously, there are two sets of precision combs, one for the chamber RO-end and the other for the HV-end. Figure~\ref{fig:oneset-comb} shows one set of the combs stacked on the granite table. The comb geometry  precision was measured three ways: comb flatness, and the horizontal and vertical pitches of the tube positions defined by the holes in the combs.

\begin{figure}[hbt]
    \centering
    \subfloat[]{
    \includegraphics[width=0.3\textwidth]{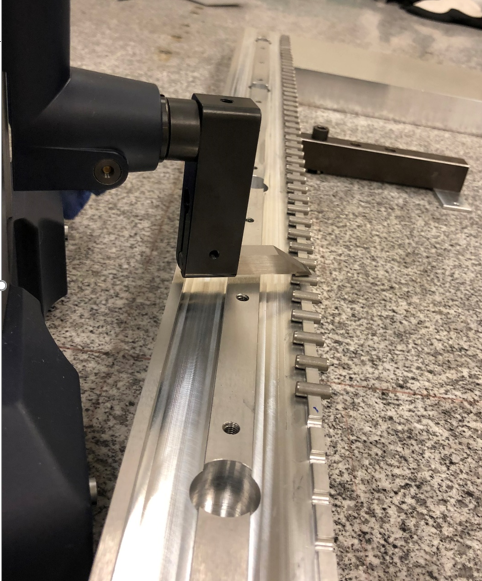}
    }
    \hspace{0.5cm}
      \subfloat[]{
        \includegraphics[width=0.24\textwidth]{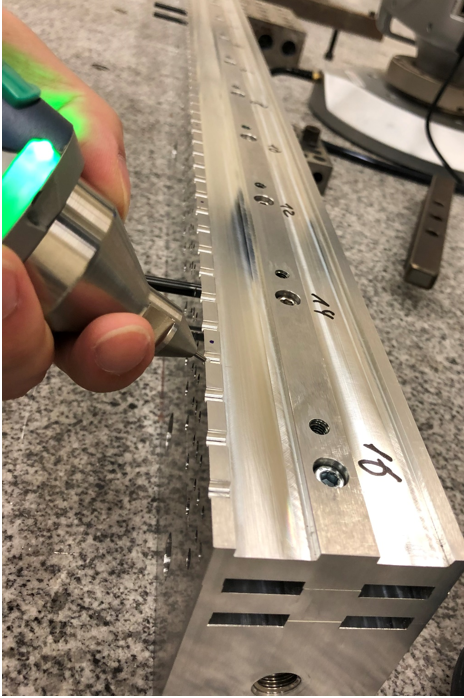}
        }
    \caption{Precision comb geometry    
            measurements: (a) flatness
            using the height gauge; 
        (b) wire pitch using the 
            FARO-arm to determine the center 
            positions of the comb half-cylinders.}
    \label{fig:combgeometry-measurements}
\end{figure}

\begin{figure} [!htb]
    \centering
    \includegraphics[width=0.8\textwidth]{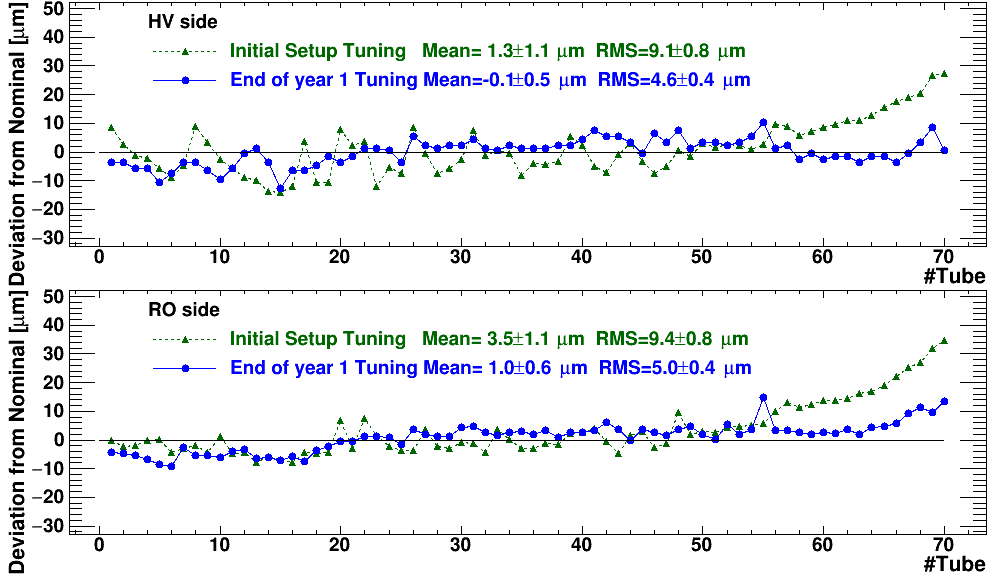}
    \caption{Base comb cylinder center height 
        measurements relative to the target 
        wire height above the granite surface 
        for HV and RO sides, showing the initial
        tuning and the re-tuning after assembly 
        of the 12th chamber.
    }
    \label{fig:base-tuning}
\end{figure}

The flatness of the base-combs was measured 
with the height gauge using precision rods placed
in the combs precision half-circles (see figure~\ref{fig:combgeometry-measurements}(a)). 
Both base combs, when first set up on the 
granite, had a bowed shape, where the ends were 
about 80 microns higher than the middle.
This same shape was seen in subsequent layers 
as they were added to the base comb.
Flattening of the base combs was achieved 
by an iterative process of loosening the 
screws holding the ends of the base to the 
side bars, applying downward pressure on the 
outer ends and re-tightening the screws 
followed by height measurements, and 
repeating the whole process until the 
combs were sufficiently flat.
The measured shapes after the flatness tuning 
of the readout (RO) and high voltage (HV) 
side base combs are shown in 
figure~\ref{fig:base-tuning} 
(green data-points).
The measured standard deviation (RMS) of the 
comb flatness is smaller than 10~$\mu$m 
for both base-combs.
The initial indexed tube locations above 60 
show higher deviations, at the level of 20 microns.
After the construction of the first twelve sMDT
chambers, the RO and HV base comb shapes were 
re-flattened using the method described above 
The blue data-points in 
figure~\ref{fig:base-tuning}. 
show and improvement of the average flatness 
to 1~$\mu$m or better and of the RMS from 
9~$\mu$m to 5~$\mu$m.
\par
\begin{figure}[!htb]
    \centering
    \includegraphics[width=0.8\textwidth]{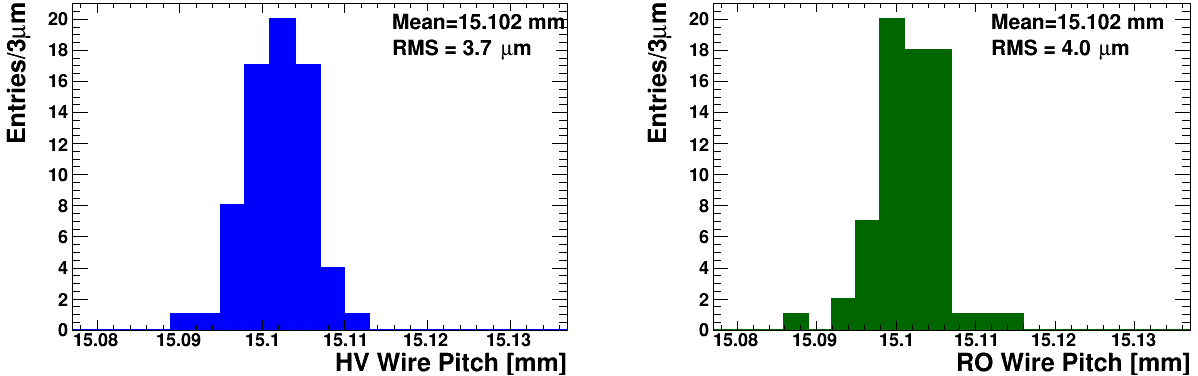}
    \caption{Measurements of the base-comb 
        wire pitch, for both HV and RO side.
    }
    \label{fig:basecomb-wire-pitch}
\end{figure}

The Faro-arm with a precision spherical probe tip
was used to measure the half-cylinder center 
positions on the combs to determine the wire 
pitch (see figure~\ref{fig:combgeometry-measurements}(b)). 
Measurements of the deviations of the base 
comb cylinder center pitch (wire pitch) 
from the nominal value (15.100~mm) 
are shown in figure~\ref{fig:basecomb-wire-pitch}. 
The horizontal wire pitch were measured to be 
$ 15.1017 \pm 0.0005 $~mm ($ 15.1019 \pm 0.0004 $~mm)
for the RO (HV) comb, with an RMS of about 4 microns.

\begin{figure}[!htb]
    \centering
    \includegraphics[width=.85\textwidth]{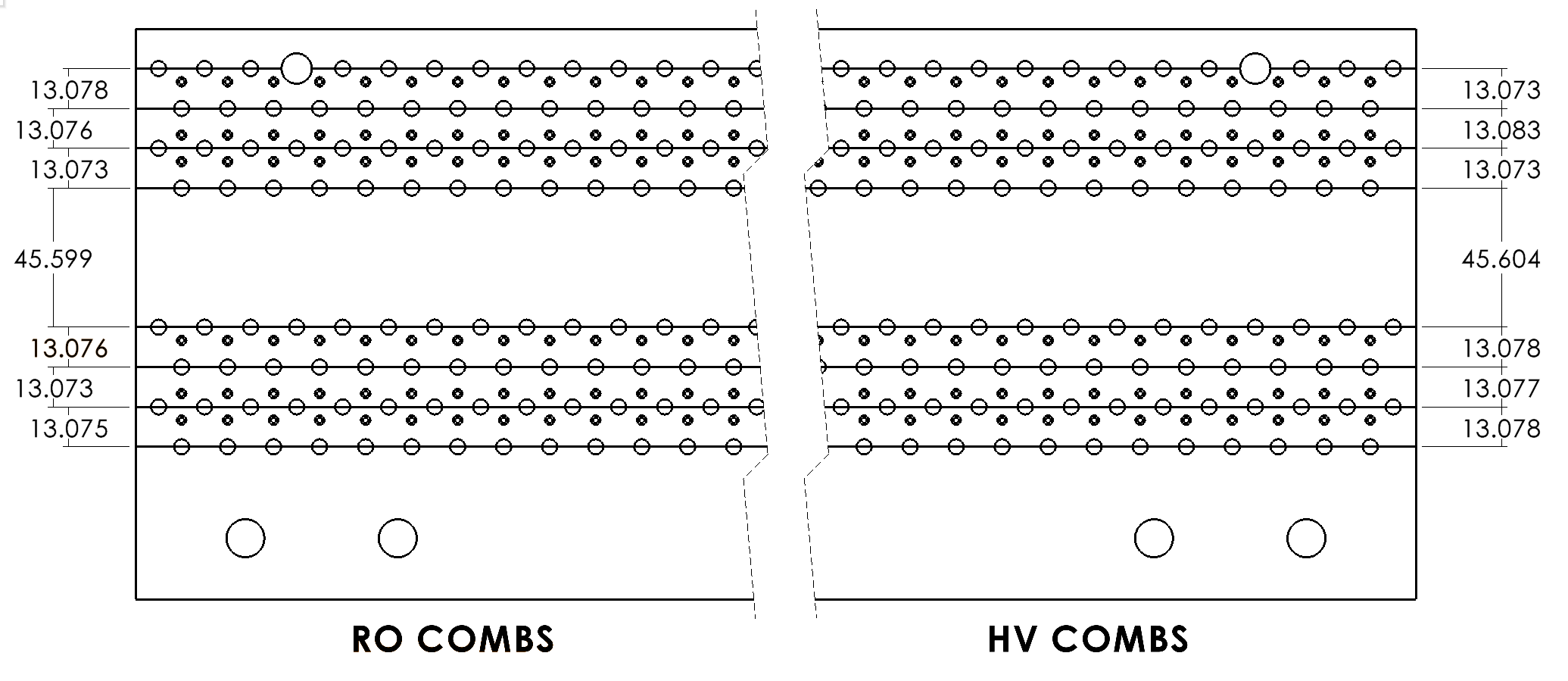}
    \caption{
    Summary of the measured precision comb vertical 
    tube-pitch, for both RO and HV side after set 
    up on granite table.  Values are in mm.  
    }
    \label{fig:Comb-geometrys}
\end{figure}

The complete geometry of the vertical-pitch 
(tube layer pitch) on both RO and HV sides was
measured on the granite table with 5 microns
precision using the height gauge  and shown in 
figure~\ref{fig:Comb-geometrys}.
The tube layer pitch was measured to be 
$ 13.077 \pm 0.005 $~mm ($13.076 \pm 0.005$~mm) 
and the spacer frame height 
$ 45.604 \pm 0.005 $~mm ($ 45.599 \pm 0.005 $~mm) 
for the HV (RO) side. 

\subsection{Spacer frame assembly jigging and 
in-plane alignment system}

\begin{figure}[!hbt]
    \centering
    \subfloat[]{
     \includegraphics[width=0.355\textwidth]{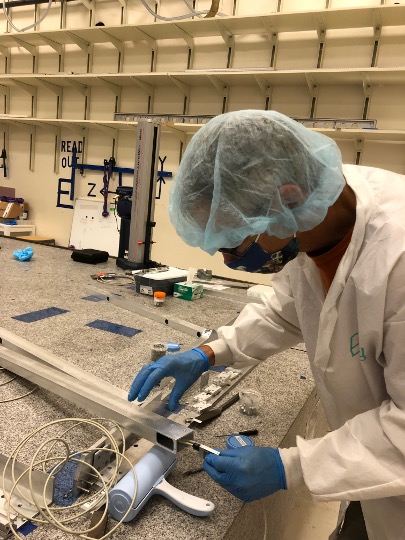}
    }
    \subfloat[]{
        \includegraphics[width=.50\textwidth]{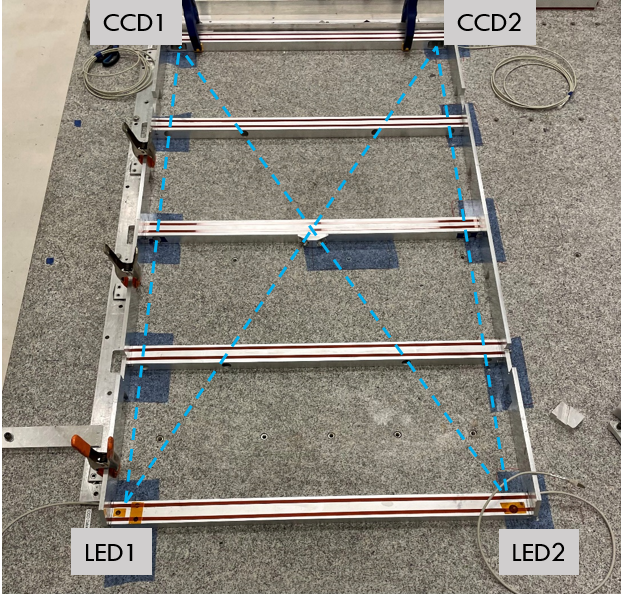}
    }
    \caption{
    (a) Assembly of the in-plane system parts into 
        the cross-bar;
    (b) spacer frame assembling with a 
        jigging setup on the granite table.
        The locations of the in-plane devices, 
        LEDs and CCDs, inside the cross bars are
        indicated as well as the four RASNIK lines, 
        (blue lines).
    }
    \label{fig:spacer frame}
\end{figure}

The spacer frame consists of five aluminum 
extrusions with rectangular cross-sections 
of dimensions 
$\rm 880 \times 30 \times 50 \; mm^{3}$.
The in-plane optical devices are housed inside 
the spacer frame cross-bars as shown in
figure~\ref{fig:spacer frame}(a).
The bars and two side support panels are 
assembled using the jigging shown in 
figure~\ref{fig:spacer frame}(b).
Precision end blocks are glued to the ends 
of the cross-bars and bolted to two side-plates 
to form a rigid frame. 
The side panels are needed to preserve the 
relative positions of the cross-bars before
the spacer frame cross-bars are glued onto 
a ML and are removed at the end chamber assembly.
\par
The spacer frame jigging consists of two 
rigid aluminum L-shaped bars firmly bolted 
to the table at right angles. 
The angle between them was set using
FARO-arm measurements.
The jigging bars are machined with cutouts to 
match the side panel profile and several L-shaped
brackets are mounted on them to reinforce the
clamping.
One side-panel and one cross-bar are clamped 
to the jigging during the assembly process to 
ensure the flatness and perfect rectangular 
shape of the frame while the glue is curing.

\begin{figure}[!hbt]
    \centering
    \includegraphics[width=0.95\textwidth]{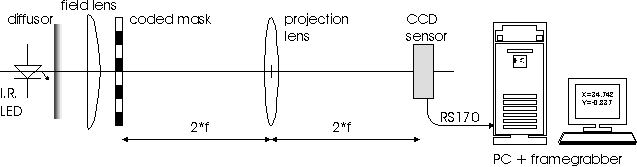}
    \caption{RASNIK system showing LED illuminating a 
    coded mask on left, imaging lens in the center, 
    and CCD on the right.
    The CCD image is read out and analysed  by a computer.}
    \label{fig:rasnikschematic}
\end{figure}

The in-plane alignment system consists of 4 RASNIK 
lines, 2 diagonal, and 2 straight, along the sides 
of the chamber. 
The RASNIK system is a three-point system 
(coded mask, lens, CCD) that is sensitive to 
the relative movement of any of the three 
elements.
A schematic of a RASNIK system with readout 
system (provided by NIKHEF and customized 
at UM as part of the chamber assembly tooling)
is shown in figure~\ref{fig:rasnikschematic}.
The coded mask is back-illuminated by an LED 
%(see figure~\ref{fig:in-plane-RO}(a))
and imaged with a lens onto a CCD.  
Analysis of the image of the coded mask 
%(in figure~\ref{fig:in-plane-RO}(b)) 
reveals displacements transverse to the surface 
of the mask accurate to 1 micron. 
\par 
The in-plane system has 2 LED-Masks and 2 CCDs.  
Each CCD views sequentially images of each 
LED-Mask to provide one straight and one 
diagonal measurement.
Reference RASNIK measurements are taken after 
chamber assembly on the granite table when
the chamber is considered to be in perfect 
alignment while still held by the precision
jigging.  
RASNIK measurements taken after the chamber 
is moved from the granite reveal chamber
deformations when subtracted from the granite
reference RASNIK measurements.

\subsection{Survey of the jigging for sensor platform installation}
\label{sec:platforms}

Two different jigging sets (one for BIS1 and 
another for BIS2-6 chambers), provided by MPI, 
are used to install the sensor platforms on top 
of each chamber while it is still on the combs, 
so that the platform locations can be set and 
measured with precision.
The combs determine the geometric grid of wires 
within a chamber. 
Knowledge of where one chamber's wires are 
relative to its neighbors in the same plane 
is provided by a set of Axial-Praxial (AP)
sensors. 
The exact locations of 3 planes of each AP 
platform relative to the wire grid must be 
accurate within 0.20~mm and known to within 
0.03~mm.
Furthermore each chamber has two Hall sensors 
(for measuring local magnetic field direction 
and intensity) and some chambers have CCC 
(Chamber to Chamber Connection) sensors for 
measuring a small (large) chamber's location 
relative to chambers in the adjacent large (small) 
sector.
The AP platforms and CCC platforms must be glued 
with X, Y and Z positions accuracy each better 
than 200~microns compared to nominal designed
positions.
The precision on the measured 3D-positions are 
each required to be within 30~microns when 
referenced to the granite surface (Y-axis), 
endplug surface (X-axis), and the first wires 
on both sides of the chamber (Z-axis). 
The B-sensor platforms must be glued with 
positions better than 500 microns compared 
to nominal designed positions.
\begin{figure} [!htb]
    \centering
    \subfloat[]{\includegraphics[width=0.62\textwidth]{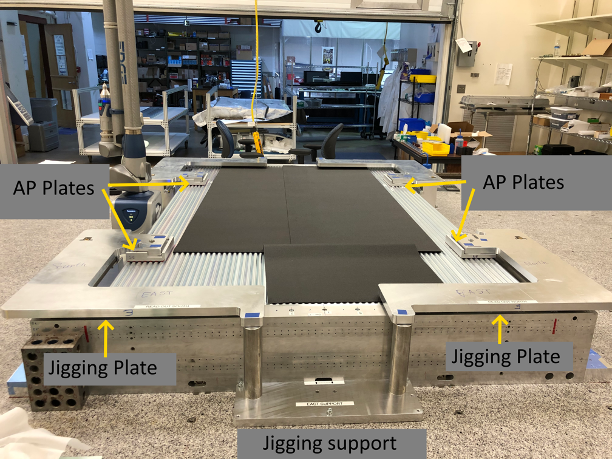}}
\hspace{0.1cm}
\subfloat[]{\includegraphics[width=0.35\textwidth]{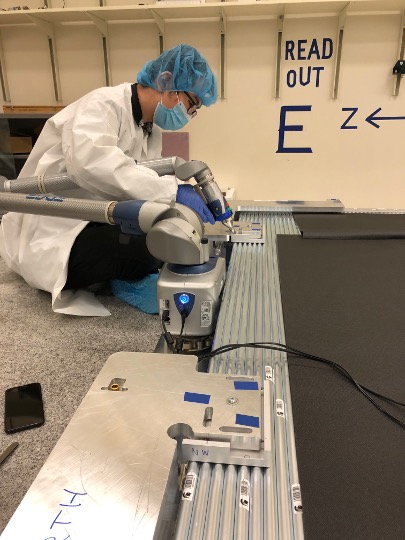}}
    \caption{(a) Four AP plates on the top 
    corners of a sMDT chamber, each held by a 
    screw to a L-shaped jigging;
    (b) a FARO-arm probe is used to check and 
    adjust locations of AP platforms before 
    and after gluing. 
    Final positions are measured after glue is cured.}
    \label{fig:Plate-measurement}
\end{figure}

The AP jigging references the comb precision
holes to set each AP platform within 40~microns
of a reference location.
The AP jigging supporting structures, two 
close to the end-combs and four mounted on 
the base cross bars, are tuned and measured 
so that all the surfaces of the L-shaped 
jigging are parallel to the surface of the
granite table.
Each L-shaped jigging 
(see figure~\ref{fig:Plate-measurement}(a)) 
holds one platform and allows fine 3D 
adjustments to be made via 6 contact points, 
two referenced to the first wire, one 
relative to the tube RO endplug surface and
three to the surface of the granite table.
Each AP-plate's position is verified by 
FARO-arm measurements as shown in 
figure~\ref{fig:Plate-measurement}(b) 
before and after gluing.
\begin{figure} [!htb]
    \centering
    \subfloat[]{
    {\includegraphics[width=.41\textwidth]{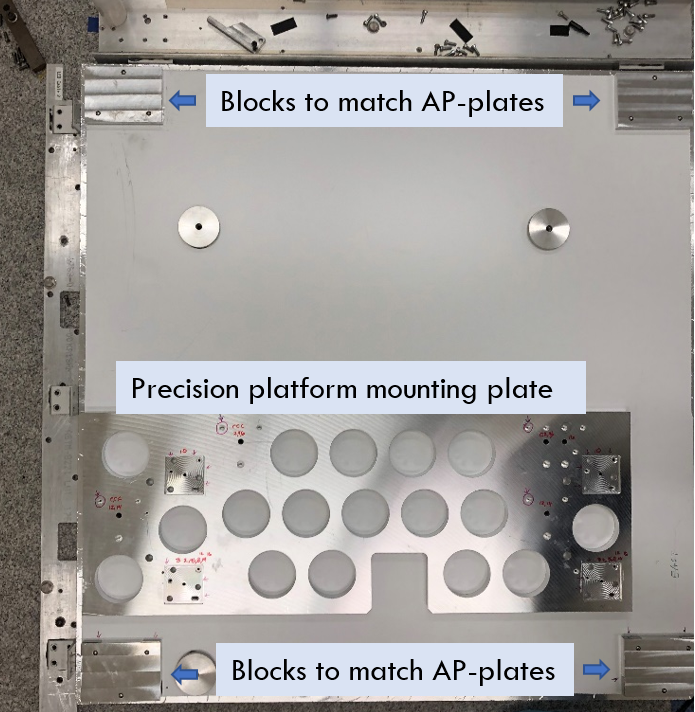}}
    }
    \subfloat[]{
    {\includegraphics[width=.565\textwidth]{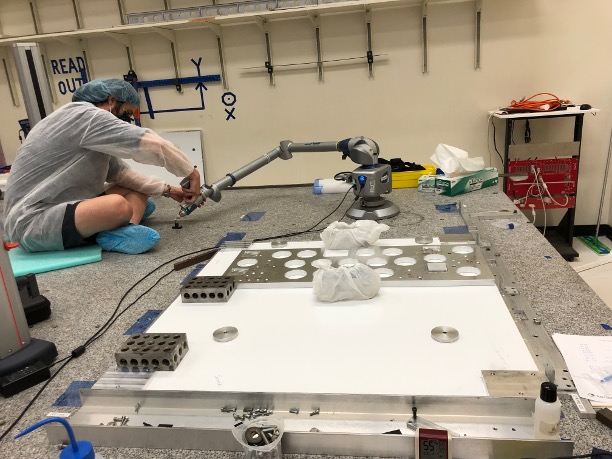}}
    }
    \caption{(a) Assembly plate to mount the
        B-field and CCC platforms for installation
        on chamber;
        (b) survey of the big plate with the
        FARO-arm.
    }
    \label{fig:platformSurvey}
\end{figure}

The CCC-plates and B-sensor plates are mounted 
on a big precision plate where their positions
are fixed relative to four blocks matching 
the AP-plate positions on the chamber
(see figure~\ref{fig:platformSurvey}(a)).
Surveys on two assembly plates (for BIS1 and 
BIS2-6 chambers) were done using the FARO-arm 
on the granite table
(see figure~\ref{fig:platformSurvey}(b)).
Mistakes were found on both of them. 
One was fixed by removing one of the four blocks 
and re-gluing it on the correct position on the
plate. 
The other plate was rectified by gluing a 
120~microns shim to the reference block which 
sets the X position of the platforms on the 
chamber, 
thus correcting the relative position of the 
platforms respect to the block.
All the fixes were further verified by dry runs 
of installation of the platforms on a chamber.

\subsection{Chamber test stations}

Two chamber test stations were designed and built 
at UM.
One is used for installation of on-chamber 
services (temperature sensors, cables, on-chamber 
gas system, temperature and RASNIK readout panel, 
HV and RO  electronics hedgehog 
cards\footnote{Hedgehog cards are front-end electronics cards which attach directly to the 
tube endplugs.}, 
and Faraday cages) and for performing a gas
tightness test.
The other test station is dedicated to 
electronics and cosmic ray tests.

\subsubsection{Chamber service installation and leak test station}

\begin{figure} [!htb]
    \centering
    \subfloat[]{
        \includegraphics[width=.49\textwidth]{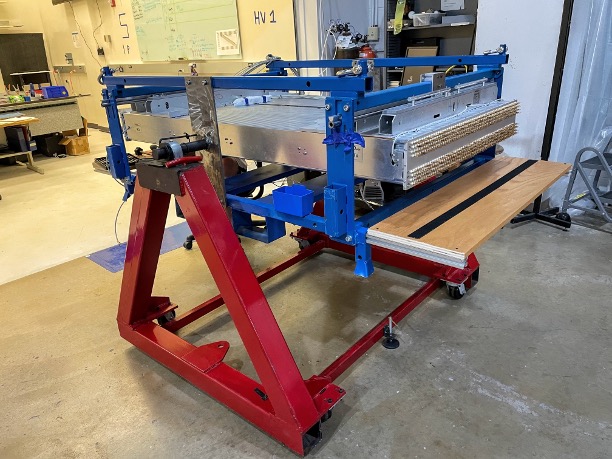}
    }
    \subfloat[]{
    \includegraphics[width=.49\textwidth]{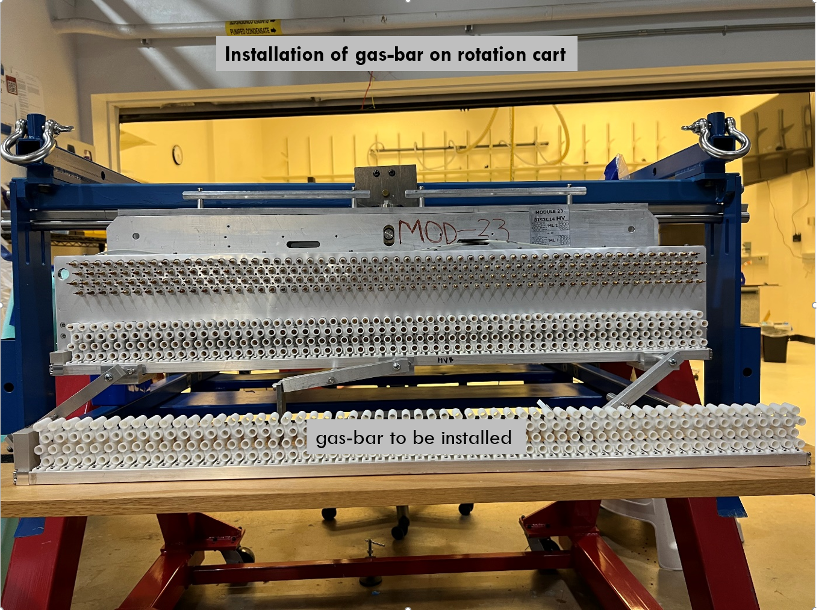}
    }
    \caption{(a) A rotation cart for chamber
        service installation and testing;
        (b) Installation of the gas-bars on
        the rotation cart.
    }
    \label{fig:rotation}
\end{figure}

The chamber services installation is done 
by mounting the chamber on a rotation cart 
consisting of a chamber-frame (provided by
Protvino) mounted on rotation bearings on a 
movable base (figure~\ref{fig:rotation} (a)).  
The chamber is attached to the frame using 
the chamber kinematic mounts. 
On the cart the chamber can be rotated through 
$360^{\circ}$ on the base allowing easy access 
of all parts of the chamber, for example, when
gluing  temperature cables on top and bottom 
of the  chamber. 
Two removable small tables can be connected to 
the chamber-frame for services installation 
tooling and parts. 
The rotation feature is also used to do a test 
of the in-plane alignment system by taking RASNIK 
measurements at different rotation angles and 
checking for a characteristic set of deflection 
measurements.
Chamber ground cables, front-end hedgehog cards 
and Faraday cage installations on chambers 
are carried out in this station.
An end-view of a chamber mounted on the rotation 
cart is  shown in figure~\ref{fig:rotation}(b),
where one gas-bar is installed on the bottom
ML of the chamber, and one gas-bar is on the 
small table ready to be installed.

\subsubsection{Cosmic ray test station}
\label{sec:cosmicTestStation}

\begin{figure}[!hbt]
    \centering
    \includegraphics[width=0.8\textwidth]{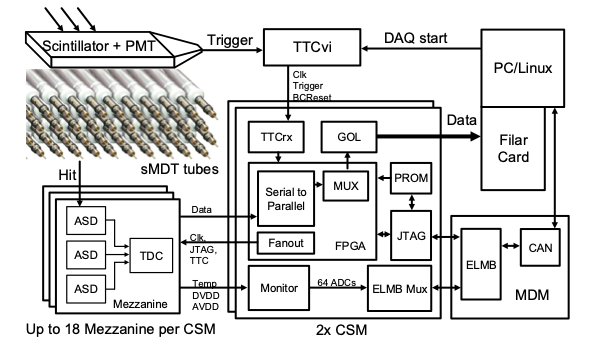}
    \caption{Block diagram of the sMDT readout electronics and the DAQ system.}
    \label{fig:miniDAQ}
\end{figure}

The chamber cosmic ray testing is performed inside 
a room where humidity is kept below 40\%.
The station itself consists of a scintillator 
trigger system mounted in a movable frame, 
a chamber cart with a gas control panel,
HV/LV power supplies, and a data acquisition 
(DAQ) system~\cite{TDCandMiniDAQ}.  
The schematics of the cosmic ray test system, 
including an sMDT chamber, trigger detector, 
front-end electronics, the DAQ modules and 
data flow, is shown in figure~\ref{fig:miniDAQ}.

\begin{figure} [!htb]
    \centering
        \subfloat[]{
{\includegraphics[width=.45\textwidth]{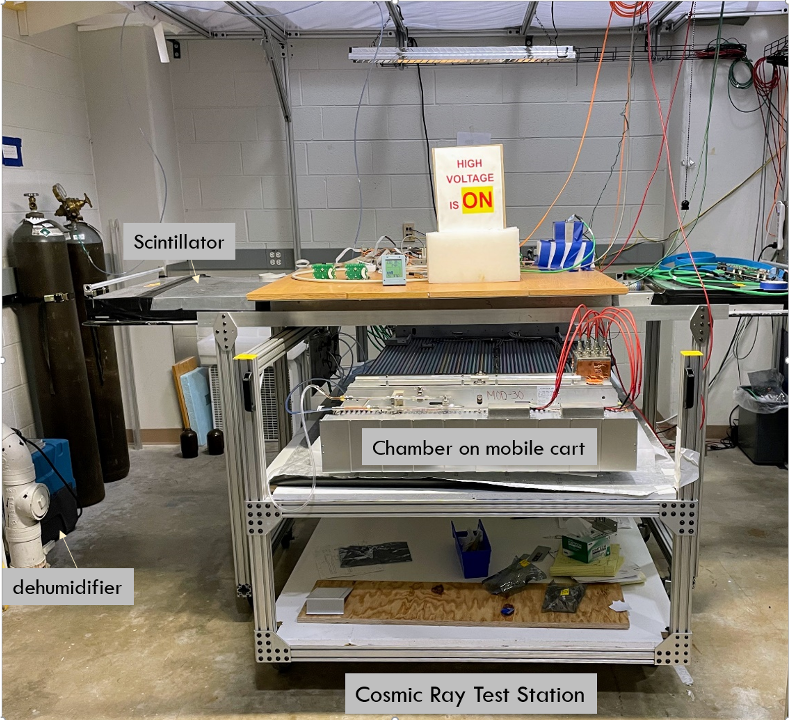}}
}
 \subfloat[]{
{\includegraphics[width=.55\textwidth]{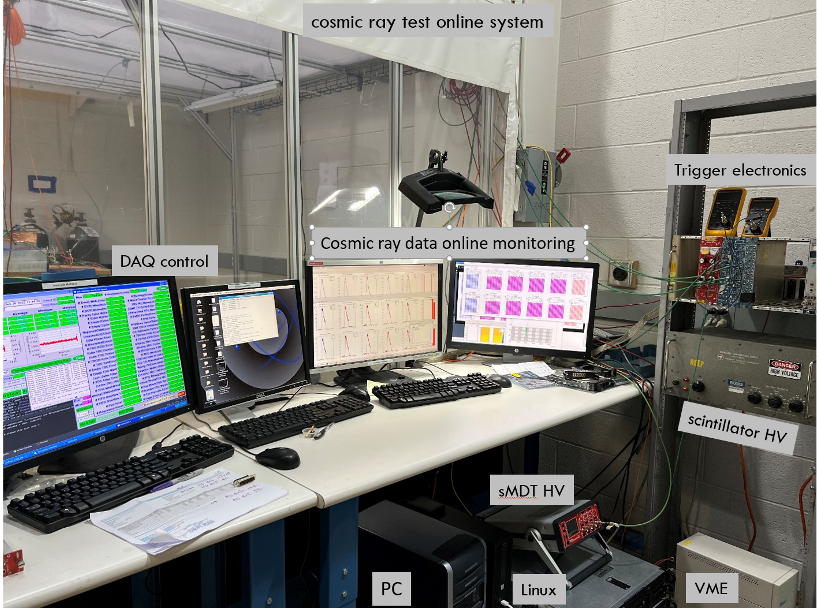}}
}
    \caption{%Cosmic ray test station: 
    (a) Scintillation pad on a mobile support
        over a test chamber on a mobile
        cart connected with gas and HV;
    (b) online DAQ and monitoring systems for
        the cosmic ray test.
    }
    \label{fig:cosmic-station}
\end{figure}

Figure~\ref{fig:cosmic-station} (a) shows an 
sMDT chamber on the mobile cart with the scintillator
($1.2\times 0.6~m^2$) placed on top 
to provide fast trigger signal
for the cosmic ray test.

The DAQ system (see
figure~\ref{fig:cosmic-station}(b)) is located 
outside the humidity-controlled chamber test room 
and consists of two computers (a Windows computer
for DAQ configuration and controls and a linux 
computer for data-taking), a VME crate containing
the DAQ interface and trigger modules, HV power 
supplies for chamber and for the trigger 
scintillator, and a NIM crate containing the 
trigger coincidence modules. 
The firmware of the DAQ and the software for online 
data monitoring were developed at Michigan. 
During the cosmic ray test the muon tracks, 
hit map, and the ADC and TDC spectra for each 
sMDT tube are displayed online, as shown in 
figure~\ref{fig:cosmic-station}(b).

\section{sMDT Base Chamber Construction on Granite Table}
\label{sec:construction}
This section describes the assembly of the base 
chamber  on the granite table using the jigging 
described in Section~\ref{sec:infrastucture} as 
well as measurements of the chamber mechanical 
precision done on the granite table after chamber 
assembly.

\subsection{Spacer frame and in-plane alignment 
system assembly}
\label{sec:spacer-frame}

The spacer frame parts are first cleaned with 
IPA and then are assembled with the in-plane 
devices and tested before being glued to the
chamber.
The mechanical shape of the frame is determined 
by the assembly jigging while its flatness,
referenced to the surface of the granite table, 
is checked with 25 micron shims. 
\par
In the spacer frame the optical devices used
in the four RASNIK lines are installed 
as shown in figure~\ref{fig:spacer frame}.  
The LED+mask and CCDs are mounted on the end 
cross-bars of the spacer frame, and the lenses 
for each RASNIK line are mounted on the center
cross-bar. 
Images are taken after spacer frame is assembled 
to certify that the hardware is fully functional.
Final reference RASNIK measurements are taken 
after the chamber is fully assembled and still
held by the precision combs.  

\subsection{Gluing an sMDT base chamber}

The tubes that have passed the quality control 
tests (including straightness,  wire tension, 
gas leak, and dark-current 
measurements~\cite{UMtubePaper}) 
are cleaned with IPA before chamber gluing. 
They are arranged in eight groups by layer and 
their barcode ID is recorded in the database 
together with their final positions on the chamber.
In addition, the precision combs are also cleaned 
with IPA and the tube ground screws inserted into 
the dedicated holes in the combs.
An sMDT chamber is then assembled by gluing eight 
tube layers and a spacer frame with the precision 
jigging set up on the granite table.
This sequence takes 5 days: two days to glue the 
bottom multi-layer, one day to glue the spacer
frame  and the first tube layer of the second 
multi-layer, and the last two days to glue the
remaining three tube layers on the top 
multi-layer.
\par
\begin{figure}[htb]
    \centering
     \subfloat[]{
     \includegraphics[width=.52\textwidth]{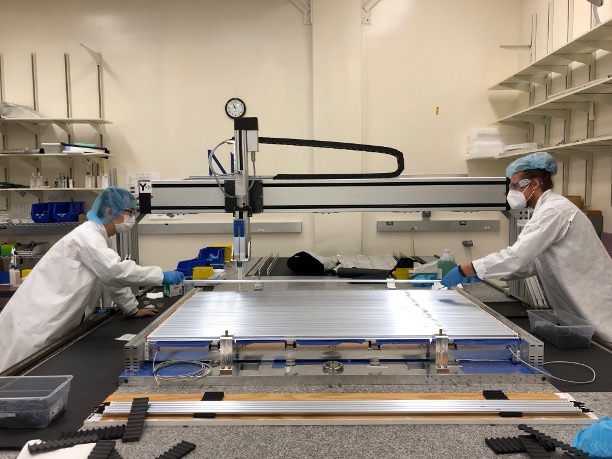}}
     \vspace{0.2cm}
    \subfloat[]{
    \includegraphics[width=.442\textwidth]{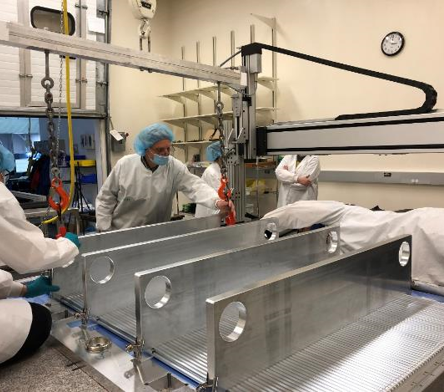}}
    \caption{(a) Two operators are placing a tube
        on the comb following the automatic glue
        dispenser on the lower layer of the tubes. 
        (b) Weights on tube layers after gluing.
    }
    \label{fig:glue-chamber}
\end{figure}

Tube layer gluing is carried out by two operators
(as in figure~\ref{fig:glue-chamber})
placing the tubes on precision combs with all 
endplug  plastic insulators adjusted to touch 
the RO-side end comb, so that tubes are lined 
up to allow proper electronics RO card 
installation ater on.
The automatic glue-machine dispenses Araldite 
2011\footnote{https://krayden.com/technical-data-sheet/huntsman-araldite-2011-technical-data-sheet/}
glue dots at 11 equidistant positions in each 
groove between the tubes. 

The pre-assembled spacer frame is glued 
between the two multi-layers (MLs) of a chamber 
using DP190\footnote{https://multimedia.3m.com/mws/media/1235570O/dp190-scotch-weld-technical-data-sheet.pdf} epoxy.
The glue is dispensed on the top of the bottom 
ML tubes in a pattern following the spacer frame 
extrusion profile. 
Four temporary combs are placed on top of the 
bottom ML to help constrain the tube 
positions when the first tube layer of the top
ML is placed on the two end combs and
glued on the spacer bars.
\par
After a layer of tubes is glued with the lower 
assembly  combs, the next level of combs are 
screwed on the top of the former one on both 
sides with a torque of 7.5~Nm for each screw.
\par
After the tubes are glued on the combs each day, 
4 weights, each with the bottom side machined to
reproduce a tube layer half-cylinder profile, 
are placed on top of the tube layer while 
supported on precision riser blocks,
as shown in figure~\ref{fig:glue-chamber}(b).
The riser blocks set both the height and the 
lateral location of the weights, and are also  
used to prevent the outermost tube from 
bowing out. 
After the glue cures overnight, the weights are
removed, and the ground-screws pre-loaded on 
comb layers 2, 3, 4, 6, 7, and 8 
(counting from the bottom) are driven into the gap 
between the tubes. 

\subsection{Installation of the platforms and the 
chamber supporting structure}
\label{sec:platform}

The ATLAS Muon spectrometer utilizes two types 
alignment sensors to determine the relative 
position of each chamber. 
The Axial-Praxial (AP) alignment system uses 
RASNIK sensors, and the CCC alignment system 
using SacCam 
sensors~\cite{phase2TDRMuonSpectrometer}.
The AP system requires four platforms on each 
chamber, and the CCC system requires 1 or 2 
platforms on selected chambers. 
Each chamber also utilizes two magnetic field 
sensors, which require two B-sensor platforms 
on each chamber.
The location of each platform must be known to 
better than 30~$\mu$m, and the platforms must 
be glued on the chamber within 200~$\mu$m 
(500~$\mu$m in the case of the B-sensor platforms)
of their target positions.
The installation process of the plates and support
structures takes two days: one day to glue the 
AP plates, and another day to glue the platforms 
for the CCC alignment and the B-field sensors, 
as well as the chamber support structures.
\par
\begin{figure} [!hbt]
    \centering
    \includegraphics[width=0.9\textwidth]{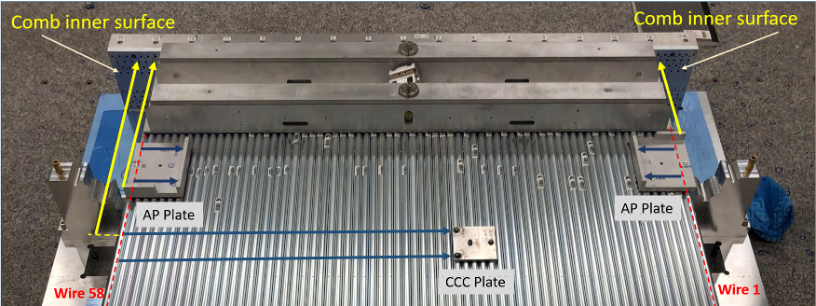}
    \caption{AP and CCC plates on top of the 
    chamber. 
    The blue and yellow arrows indicate the 
    measurement points relative to the 
    reference wires (in red) and the inner 
    RO comb surface.
    }
    \label{fig:AP-plateform}
\end{figure}

The AP-plates installation jigging was described 
in Section~\ref{sec:platforms}.
The FARO-arm is mounted on the base cross-bar of 
the chamber jigging and used to set up the 
coordinate system for the two AP-plates' 
installation on the same side. 
The FARO-arm local coordinate system is defined by 
three perpendicular surfaces: the granite table, 
the side surface of the base cross bar, and the 
outside surface of the RO side end comb.
The normal directions to these three planes 
form the coordinate axes (X, Y, Z). 
The intersection point of three surfaces determine 
the origin of the FARO-arm coordinate system.
The cylinder centers of the top layer end combs are 
measured using FARO-arm to define the line of the 
first wire in this coordinate system.
The FARO-arm measures then the 3D distances
of an AP plate relative to the tube RO endplug
surface (X), to the granite surface (Y), and
to the first (or last) wire (Z), shown in 
figure~\ref{fig:AP-plateform}.

Placing the AP-plates in correct positions is 
critical, since the assembly plate jigging 
(described in Section~\ref{sec:platforms})
relies on the AP plates to set the position 
of for B-field and CCC platforms.
Multiple measurements with the FARO-arm and
adjustments of the set screws are made
for each AP plate before gluing to ensure 
the AP plate position is within the tolerances.
DP490\footnote{https://multimedia.3m.com/mws/media/82790O/dp490-scotch-weld-tm-adhesive.pdf} epoxy
is applied to the top layer of the tubes to bind 
each AP-plate, then a new set of measurements 
is taken and further adjustment of the AP plate
position is performed if needed.
The final position of the AP plates is measured
after the glue has cured.
\par
\begin{figure} [!htb]
    \centering
        \subfloat[]{
    \includegraphics[width=0.185\textwidth]{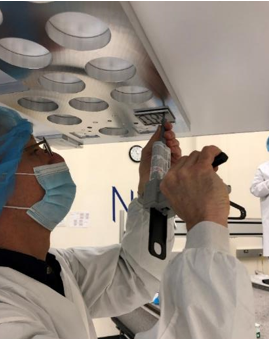}
    }
        \subfloat[]{
        \includegraphics[width=0.315\textwidth]{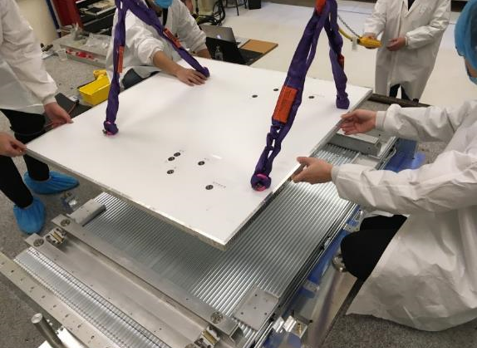}
    }
        \subfloat[]{
    \includegraphics[width=0.45\textwidth]{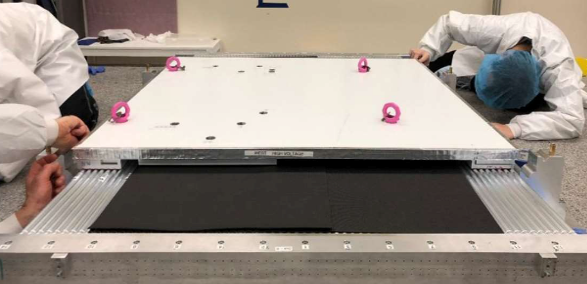}
    }
    \caption{(a) Glue being applied to B-field 
        and CCC-plates on the assembly plate;
    (b) assembly plate being placed on the AP-plates;
    (c) check of the assembly plate alignment with 
        the AP-plates on chamber.
    }
    \label{fig:big-plate}
\end{figure}

The CCC and B-sensor plates are mounted on the 
assembly plate and glued on to the chamber after 
the AP-plate glue is set.
Figure~\ref{fig:big-plate}(a) shows applying glue 
on a B-sensor plate which is mounted on the 
assembly plate prior to gluing on chamber.
Figure~\ref{fig:big-plate}(b) and (c) show lowering 
the assembly plate on top of the chamber, and 
checking that the assembly plate blocks are correctly 
positioned on the AP-plates.
\par
The chamber support structures 
(one can be seen in figure~\ref{fig:AP-plateform} 
clearly) contain the rail support bearings. 
Three kinematic mounting points, required for 
the chamber installation in the ATLAS experiment,
are located in these structures. 
A 150 micron thick insulating tape is applied 
on the bottom side of the two support structures
after they have been cleaned using IPA.
DP190 glue is dispensed by the automatic glue 
machine on to the top of the tube layer matching 
the position and profile of each support structure.
Two gauge blocks set the position of each support 
structure relative to the end-comb positions, 
and one common gauge block is used to set 
the position of the ends of the support structure 
relative to the first and last tubes.

\subsection{Measurements of the base chamber 
mechanical precision}

The chamber mechanical precision is measured 
by the 3D positions of the tubes and platforms,
using the height gauge and the FARO-arm on the
granite table, respectively.
The platform positions are first measured while 
the chamber is still held by the precision combs.
Next, the precision combs are removed, except 
the base combs and the tube positions are 
measured with the height gauge.
The in-plane system RASNIK lines are also read 
out to record the base-line measurements of the
chamber shape.

\subsubsection{Platform position measurements} 
\label{faromeasurement}

The platform positions are measured by the 
FARO-arm using the coordinates described in
Section~\ref{sec:platform}.
Several distances corresponding to reference 
points, as indicated by the arrows shown in 
figure~\ref{fig:AP-plateform}, 
are measured on the three perpendicular surfaces
of each AP and B-field platform to determine 
their 3D positions relative to the reference
positions.
Each FARO-arm measured distance is compared with 
its nominal value and the deviation from the 
target value is calculated.
\par
\begin{figure} [!htb]
    \centering
    \includegraphics[width=0.95\textwidth]{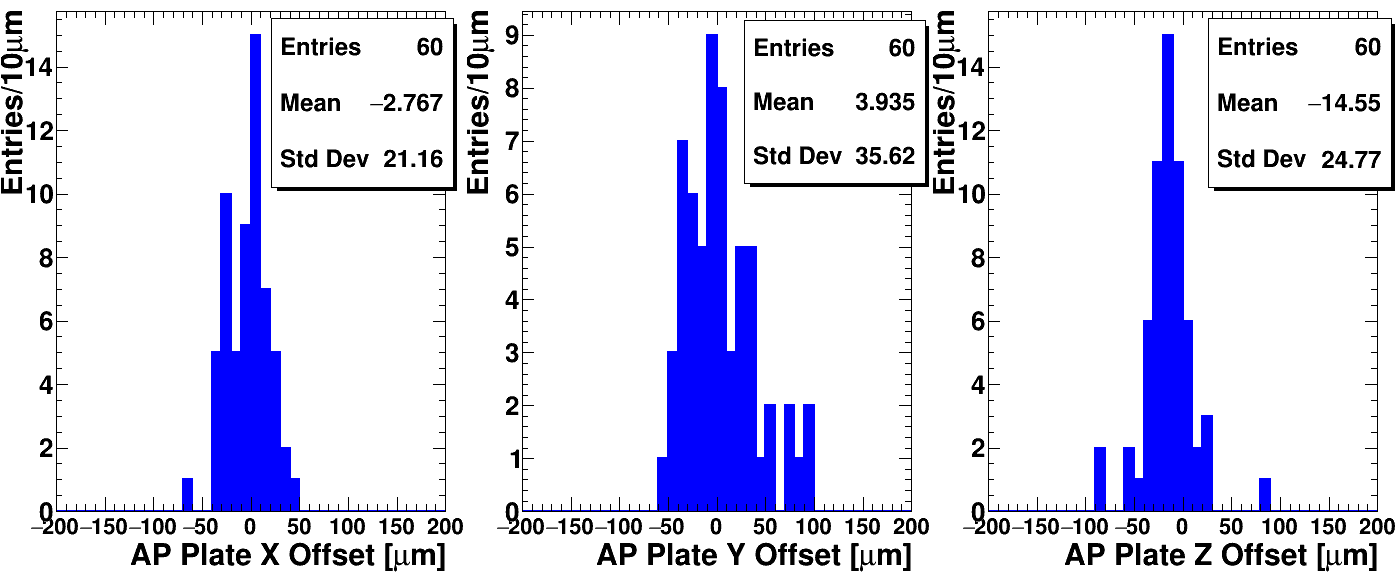}
    \caption{RO and HV AP plate position offset 
    in X, Y, and Z from target values for the 
    first 30 UM sMDT chambers.}
    \label{fig:AP-plate}
\end{figure}

Figure~\ref{fig:AP-plate} shows the measured 
AP-plate position offsets (in X, Y, Z) from the
targets for the first 30 sMDT chambers built 
at UM.
All AP-plates are placed well within the  
specified 200~microns maximum deviation 
from the nominal position. 
Similar results are obtained for B-field sensor 
positions, where the maximum tolerance is 
500~microns.

\begin{figure} [!hbt]
    \centering
    \subfloat[]{
    \includegraphics[width=0.38\textwidth]{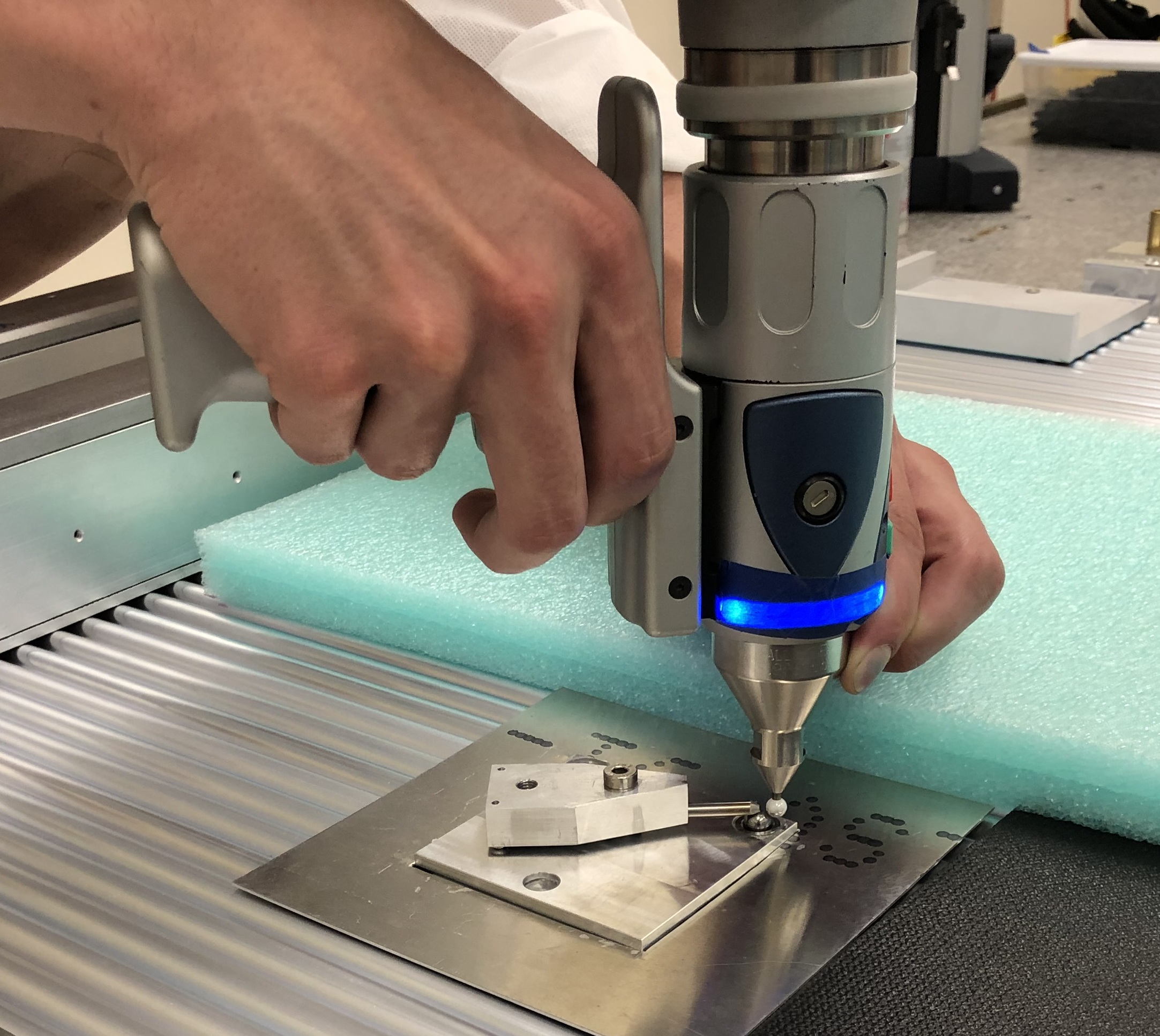}
    }
     \subfloat[]{
     \includegraphics[width=0.6\textwidth]{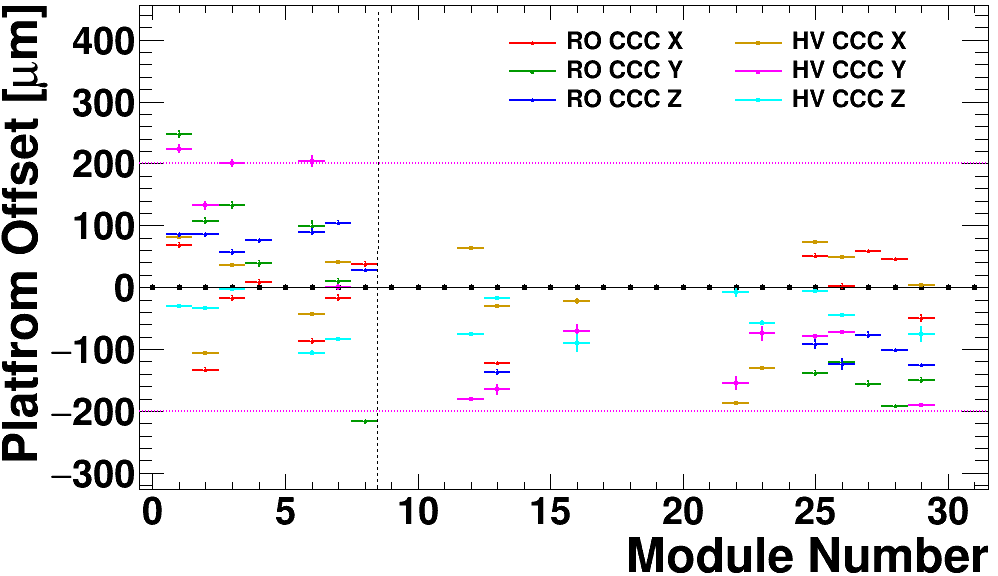}
     }
    \caption{(a) 
        A small clamping arm was devised to hold 
        the sphere rigidly while the CCC plate is 
        measured with the FARO-arm spherical probe;
    (b) Measured X, Y, and Z offsets of the CCC plate 
        positions relative to target values 
        for the first 30 UM sMDT chambers.
        The vertical black line indicates when the
        assembly plate was changed.
    }
    \label{fig:FAROMeasurement}
\end{figure}

In the case of the CCC-plates, an 8~mm precision 
sphere in a precision conical socket is measured 
to determine the center position of the plate.
A small clamping arm, as shown in
figure~\ref{fig:FAROMeasurement}(a),
was designed to hold the sphere rigidly allowing 
the FARO-arm probe to sample the surface of the 
sphere to perform this measurement.
Figure~\ref{fig:FAROMeasurement}(b) plots the
deviation of the CCC sphere center X, Y, Z 
from to the target positions.
Note that, by design, not all the chambers are 
instrumented with CCC platforms and installation
jigging was changed after building the first 
8 BIS1 chambers.
All CCC plates were installed at the target 
positions within the required accuracy of 
200~microns, except for the Y-coordinates 
for a few early chambers which were out of 
the tolerance by a maximum of 47~$\mu$m.

\subsubsection{Tube position measurements} \label{EndplugHeightMeasurement}

\begin{figure} [!htb]
    \centering
    \subfloat[]{
        \includegraphics[width=0.32\textwidth]{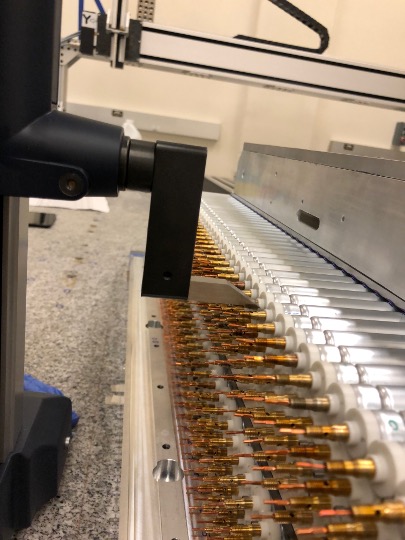}
    }
    \hspace{0.26cm}
     \subfloat[]{
    \includegraphics[width=0.58\textwidth]{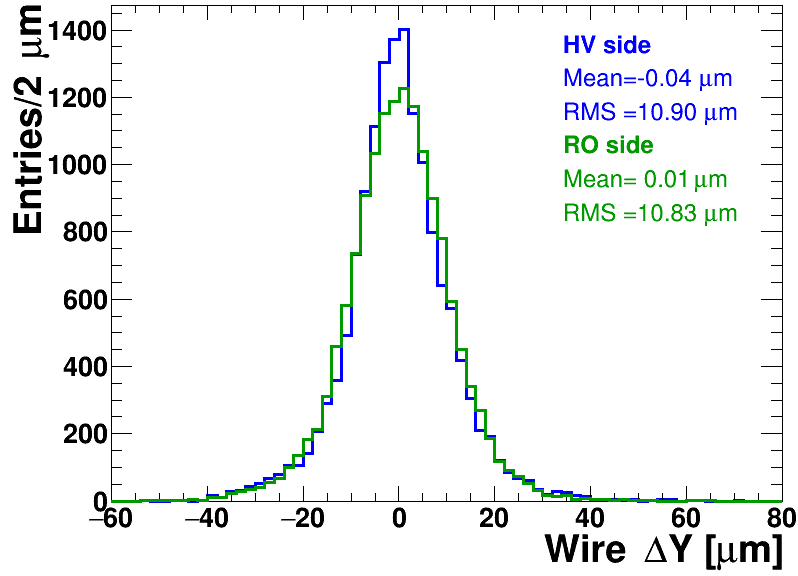}
    }
    \caption{(a) Tube height measured on the endplug
        precision cylinders with a height gauge.
    (b) Histograms of the wire height offset
        from the layer mean for the first 30 UM 
        sMDT chambers.}
    \label{fig:HVRO-tube-height}
\end{figure}

The tube height relative to the granite table 
surface is measured via the precision endplug 
cylinder surfaces using the height gauge with 
a flat probe, as shown in 
figure~\ref{fig:HVRO-tube-height}(a).
Before the measurements the granite surface used
for the measurement is cleaned. 
The flat probe is frequently checked and tuned 
with a precision block to be parallel to the
granite table.
The zero of the height gauge is set before 
starting the measurement at a specific location 
on the granite surface and re-checked if an 
abnormal value is measured.
\par
\begin{figure} [!hbt]
    \centering
    \includegraphics[width=0.49\textwidth]{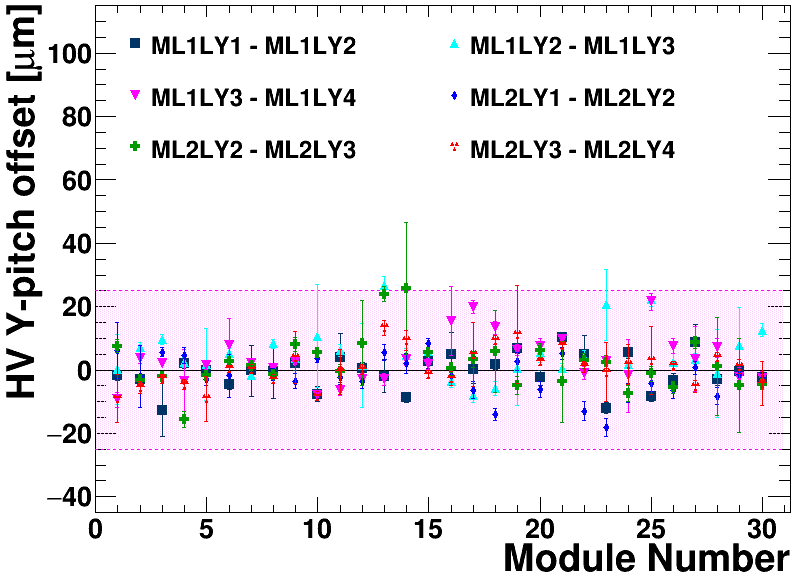}
    \includegraphics[width=0.49\textwidth]{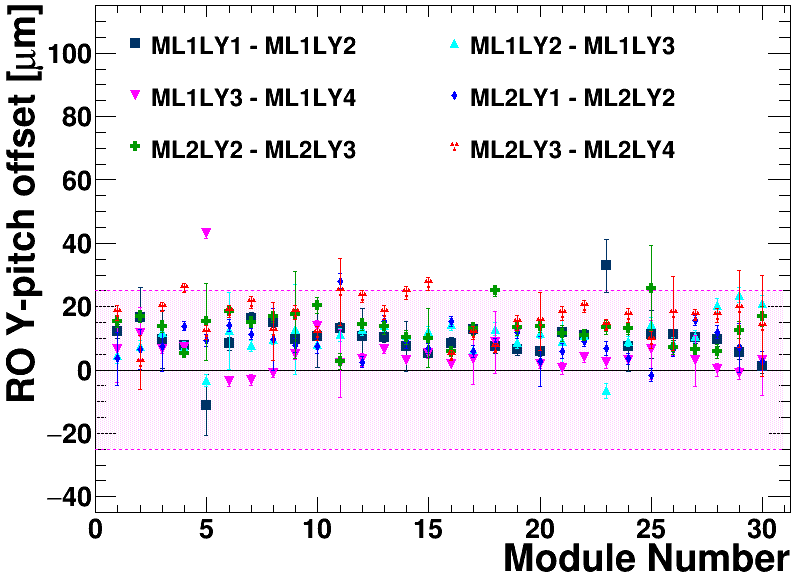}
    \caption{Tube layer pitch offset
        respect to the nominal value (13.077~cm)
        measured on both RO and HV sides for the
        first 30 sMDT chambers built at UM. 
        A black zero line and a pink band of 
        ${}\pm 25$~microns are added to guide
        the eye.
    }
    \label{fig:tube-pitch}
\end{figure}

The histograms in 
figure~\ref{fig:HVRO-tube-height}(b)
show the tube height deviations from 
the average in each tube layer for both the HV 
and RO sides.
The RMS of these distributions is below 
11~microns indicating the level of flatness 
of all the tube layers.
The average height of each tube layer is 
recorded for both RO and HV sides of the chamber,
from which the Y-pitch (vertical) of the tube 
layers and the spacer frame are determined.
Figure~\ref{fig:tube-pitch} shows all the 
tube layer Y-pitch deviation from the nominal
value for each built module. 

\begin{figure} [!hbt]
    \centering
    \includegraphics[width=0.9\textwidth]{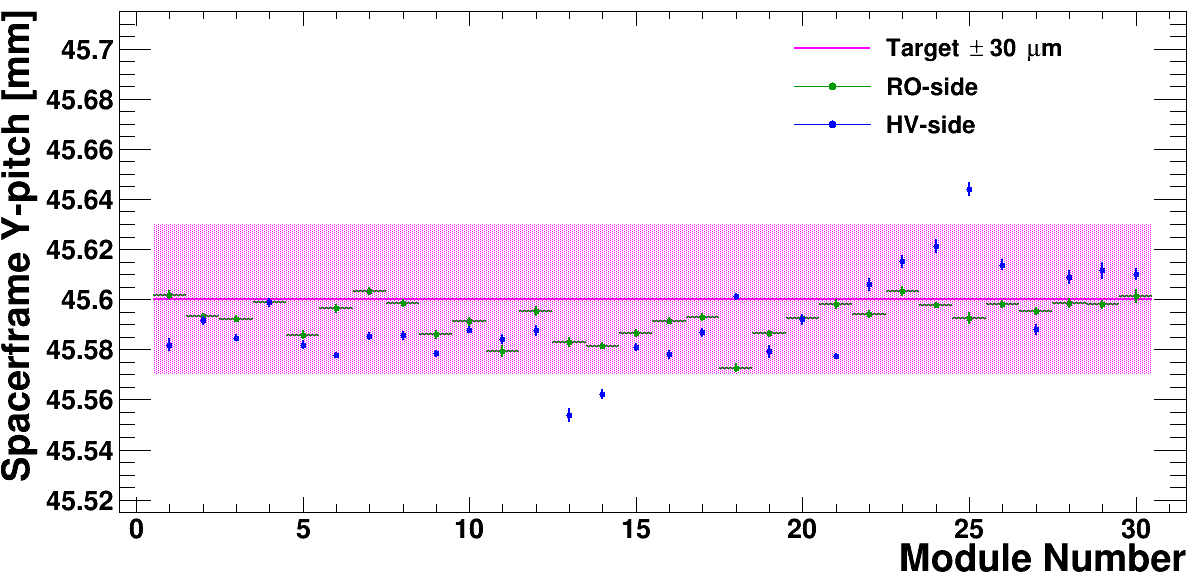}
    \caption{Spacer frame Y-pitch measurements
    for both RO and HV side for the first 30 sMDTs
    built at UM. The pink line is at the 
    nominal value of 45.600~mm.}
    \label{fig:spacerframe-pitch}
\end{figure}

The specified distance between the two 
multi-layers is 45.600~mm.
The distribution of the spacer frame distances 
is shown in figure~\ref{fig:spacerframe-pitch}.
The separation between the two multi-layers 
was found to be systematically lower than 
the target value for the first 20 chambers.
Subsequent chambers were given an additional 
50 micron mylar tape on the spacer frame 
cross-bars to increase the spacer frame 
distance to match the specification.

\subsubsection{Chamber shape deformation measurements} 
\label{sec:RASNIKrotation}

The reference RASNIK measurement is made at the 
end of the construction while the chamber is 
still in the jigging on the granite table.  
The chamber is then removed from the granite 
table and mounted on a rotation cart.
On the rotation cart RASNIK measurements are taken 
at $22.5^{\circ}$ intervals from which the granite 
reference values are subtracted to show the 
deformation of the chamber as it is rotated through 
$360^{\circ}$. 

\begin{figure}[htb]
    \centering
    \includegraphics[width=0.85\textwidth]{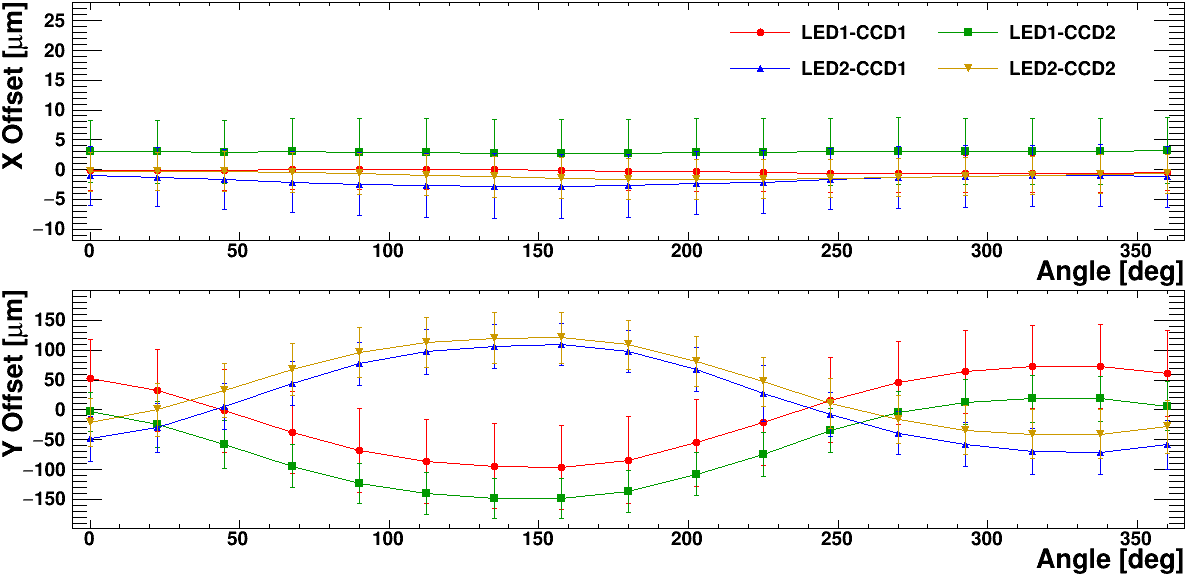}
    \caption{RASNIK measurements in chamber 
        coordinates X (top) and Y (bottom) for all
        four RASNIK lines as the chamber is rotated
        through $360^{\circ}$.
        The data points value is the averages over the
        first 30 modules built at UM, while the error
        is the standard deviation of these 
        measurements.}
    \label{fig:rasnikrotation}
\end{figure}
Figure~\ref{fig:rasnikrotation} shows the result 
of these measurements for the first 30 sMDT
chambers as deformation in the X and Y 
directions for all four LED-CCD RASNIK lines.
The X direction is transverse to the layer of 
tubes (local XZ-plane) and the chamber is very
stiff along this axis, hence there is very 
little deflection (a few microns) over
$360^{\circ}$. 
The Y direction is perpendicular to the planes 
of tubes, and much more affected by gravity, 
hence the deflection is larger with a typical 
peak-to-peak variation of  roughly 0.2~mm.
These results, besides confirming the 
functionality of the system after the cabling 
to the the readout  panel, show that the optical 
in-plane devices cover a range
of deformation of several hundredths of a 
millimeter.

\section{Chamber Services Installation and Testing}
\label{sec:services}
The chamber services are installed when the 
chamber is on the rotation cart.
The services installed are:
temperature sensors, 
ground foil, 
gas manifolds, 
hedgehog electronics cards, Faraday cages, 
and ground straps.
As soon as the chamber is removed from the 
jigging all endplugs are covered with protective
caps to keep dust from entering the tubes.
\par
Temperature sensors, together with their readout
cables, are glued in twelve positions on the
chamber surface: six on the bottom surface 
and six on the top tube layer. 
Thermal contact between temperature sensors 
and tubes is made with thermal paste epoxy.
All temperature sensor cables are routed to the 
top surface of the chamber through a hole in 
the support structure at the RO side, 
seen in figure~\ref{fig:sMDT-draw}(a).
The RASNIK cables for in-plane system control 
and readout are also routed along the top 
surface. 
A readout panel is built and installed
on the RO side support structure of the chamber
for connection to both the temperature sensor 
cables and RASNIK cables.
A perforated foil is wrapped over each end 
of the chamber, providing common ground 
connections from the tube walls to each RO 
and HV hedgehog  card. 
This foil becomes the internal wall of the 
Faraday cage that contains the electronics.
Ground pins are screwed to the grounding 
anchors (installed during the gluing operations) 
and tightened up against this foil. 
These pins are installed after installation of 
the gas system, and before the hedgehog cards 
are installed.

\subsection{Gas-bar assembly, cleaning, and 
installation on chamber}

\begin{figure}[hbt]
    \centering
    \subfloat[]{
    \includegraphics[width=0.23\textwidth]{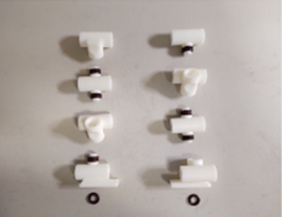}
    }
    \subfloat[]{
    \includegraphics[width=0.28\textwidth]{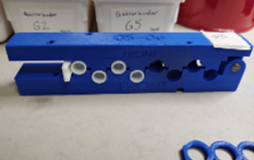}
    }
    \subfloat[]{
    \includegraphics[width=0.28\textwidth]{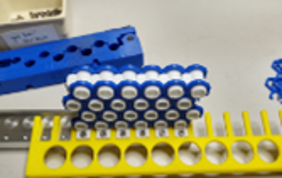}
    }
    \subfloat[]{
    \includegraphics[width=0.18\textwidth]{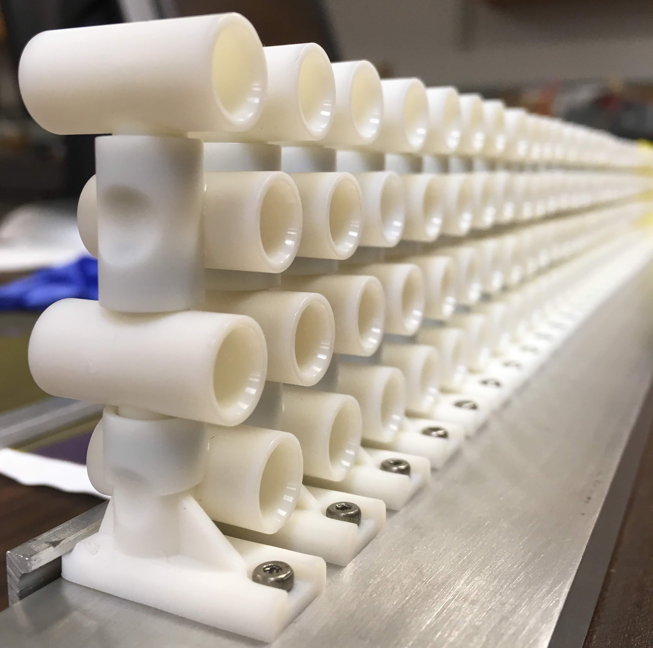}
    }
    \caption{Gas-bar components and assembly:
    (a) Molded plastic gas fittings with
        o-rings form stacks of 4 (two 
        different kinds);
    (b) blue 3-D printed jigging pre-aligns
        the 4 pieces of a stack;
    (c) stacks being mounted on a gas-bar 
        manifold using more 3-D printed
        jigging parts to aid in alignment;
    (d) a gas-bar manifold with installed
        stacks, ready to be mounted on a 
        chamber.
    }
    \label{fig:gasbar-assembly-1}
\end{figure}

The sMDT gas distribution system uses molded 
plastic parts, shown in 
figure~\ref{fig:gasbar-assembly-1}(a), 
which plug together in sets of 4 (to feed in 
series 4 tubes in a vertical stack across 
the 4 tube layers in a ML).
On both sides of a ML either 58 or 70 of such 
stacks are connected to two gas-bars for
parallel gas flow in and out of each ML. 
Gas input and output are at opposite corners 
to balance the impedance for uniform gas 
flow through all tubes.

All gas fittings are  first ultrasonically 
cleaned with IPA, which is the only lubricant
used during assembly. 
Each gas-bar is also ultrasonically cleaned 
in a bath of de-ionized water. 
After drying, each bar is wiped down with IPA. 
The inside length of the bar is then cleaned 
out using air pressure.
Each bar is again wiped down with IPA. 

Each stack of 4 fittings is assembled using 
double  o-rings between components.
A sandwich jig, shown in 
figure~\ref{fig:gasbar-assembly-1}(b)
is used to align components for assembly. 
% are mounted to the manifold gas-bar 
%using a single small screw.
The sandwich jig and several other tools 
were made with a 3D printer to simplify 
gas-bar assembly
%jigging to keep the many components properly aligned
%before installation on the chamber.
(see figure~\ref{fig:gasbar-assembly-1}(c)). 
Gas stacks are attached with M2.5X6 screws to 
the gas-bar with an o-ring in between each 
stack and the gas manifold.
Different gas-bar configurations are made 
for the two sides of the chamber.
%Once all stacks are assembled on the gas-bar, 
%as shown in figure~\ref{fig:gasbar-assembly-1}(d), 
%it is labeled with its installation 
%designations since each side of a ML requires
%a specific configuration of the gas manifold.
Completed gas-bars are stored in lint-free 
bags prior to installation on chamber.
\par
The gas-bar cleaning process was modified 
partway through chamber production because 
it was found that metal pieces left from 
the machining process could block a gas hole.
This problem was found during cosmic ray 
testing where it was seen that some sets of 
4 tubes connected to the same stack showed 
low efficiencies.  
To investigate this problem, the electronics 
and gas-bar were removed from the chamber.
Further inspection revealed that small 
circular metal chips left over from hole
drilling were blocking the holes in the 
gas-bar.
The affected chambers had to have electronics 
and gas-bar removed and the gas-bar carefully 
inspected for chips and re-cleaned.
After reassembly, the tube efficiency was 
restored to normal values.
%The gas stack was removed from the gas-bar,  
%to be the cause of  these issues. 
%The following steps were taken to attempt 
%to repair the tube stacks.
%The Faraday frame panels, hedgehog cards,
%and grounding pins in the area of the low 
%efficiency stack were removed to re-clean 
%both gas-bars mounted %on one ML.
%The Faraday frame side panels and  fittings
%on the gas-bar-ends were also removed. 
%Then, with the stack and gas-bar unobstructed,
%The low efficiency stacks were carefully 
%removed from the gas-bars. 
%While looking through the gas-bar, 
%An unbent paperclip was poked through the 
%gas-bar hole corresponding to the inefficient 
%stack. 
%In multiple cases, the paperclip dislodged 
%a circular metal chip from the gas-bar hole. 
%about the right size to 
%The chip, which significantly impeded gas-flow to the %stack, 
%was left over from machining the hole. The chip, 
%once cleared, remained in the manifold.
%Using a plunger fitted to the size and shape of the gas-bar bore, 
%the metal chips and any other debris were pushed 
%out of the end. 
%In some cases, this required wrapping the plunger
%in a tissue wipe and wetting the wipe with IPA alcohol.
%The inefficient stack was then replaced with a 
%new one, and everything that had been taken
%out was reassembled. 
%In all cases where a metal chip was successfully 
%removed, efficiencies of the tubes associated with
%that stack returned to normal levels. 
%In cases where no chip was found, replacing the stack fixed the efficiency problems.
%The metal chips are believed to be byproducts of the gas-bar machining process.
The gas-bar cleaning procedure was subsequently
modified to include the poking of each gas-bar
hole to remove any metal fragments.
Since this procedure was adopted no further
issues with low efficiency stacks have been 
observed.
\par
\begin{figure}[hbt]
    \centering
    \subfloat[]{
    \includegraphics[width=0.27\textwidth]{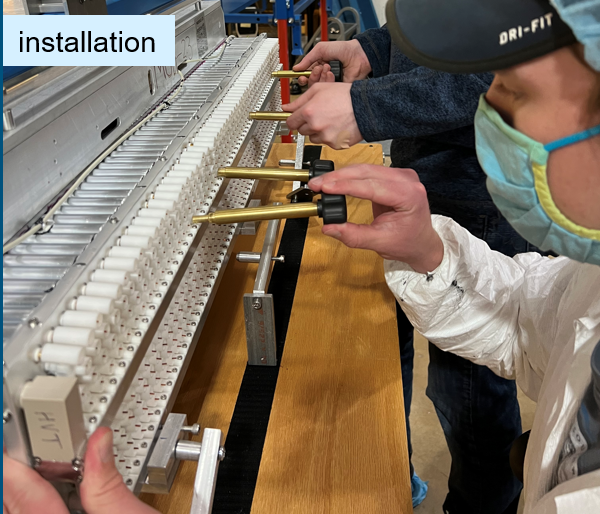}
    }
    \subfloat[]{
    \includegraphics[width=0.39\textwidth]{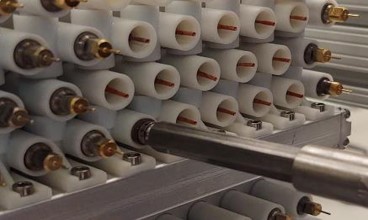}
    }
      \subfloat[]{
    \includegraphics[width=0.32\textwidth]{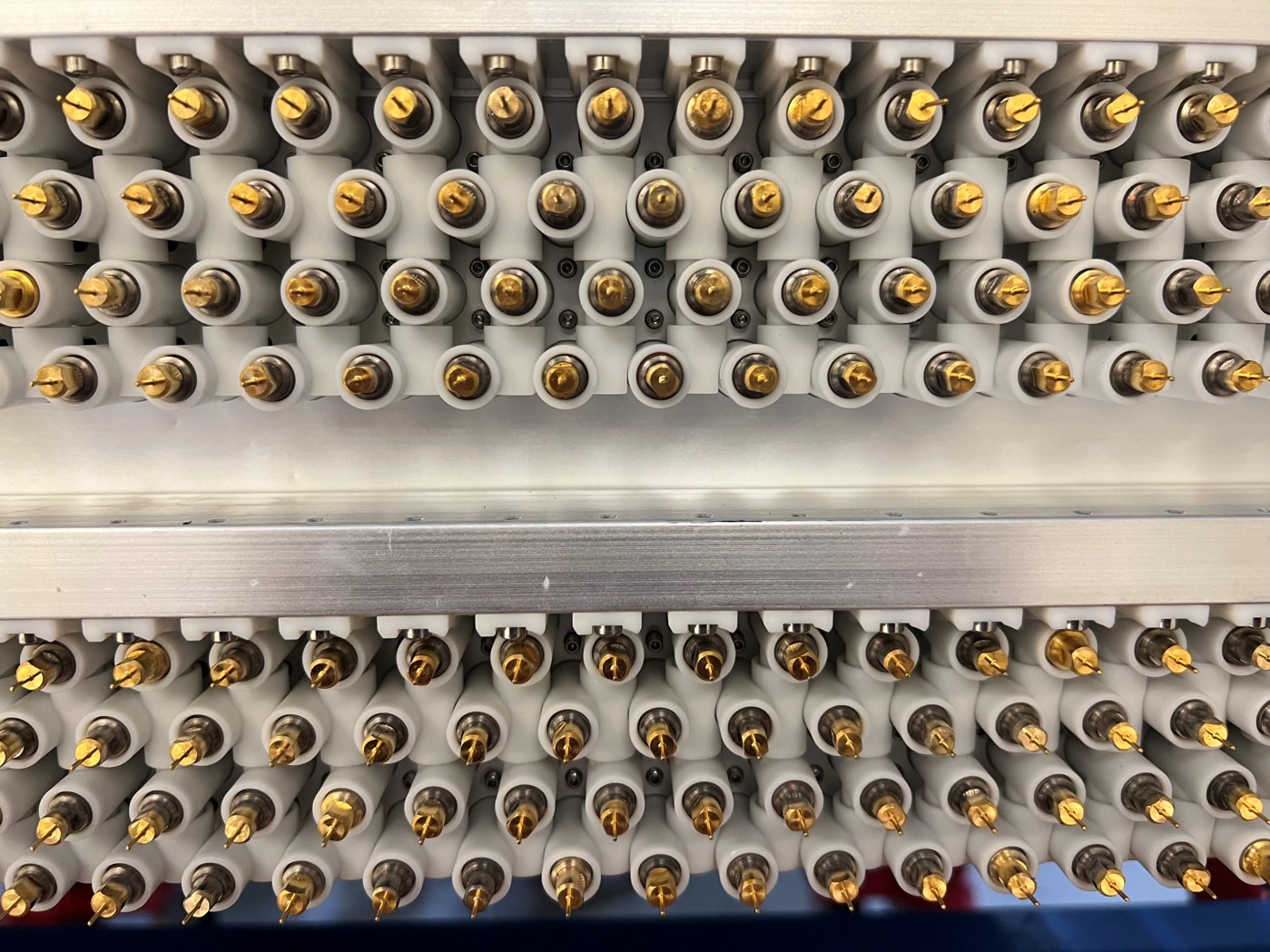}
    }
    \caption{Gas-bar installation on a chamber: 
    (a) fitting a pre-assembled gas-bar on 
        the top  ML of a chamber.
        The brass tools are used to make sure
        that all gas fittings are fully seated 
        on their corresponding endplugs before 
        beginning to install o-rings.
        The same tools are used to force an 
        o-ring over the endplug precision 
        surface where it is compressed by the 
        gas fitting;
    (b) after the back o-ring is in place the 
        signal-cap, which holds the outer 
        o-ring, is screwed onto the endplug. 
        The signal-cap has cross drilled holes 
        which allow the gas from the fitting to
        enter the endplug and the tube body; 
    (c) Two multi-layers with gas-bars and 
        signal-caps installed.
    }
    \label{fig:gasbar-assembly-2}
\end{figure}

Figure~\ref{fig:gasbar-assembly-2} shows 
the gas-bar installation on a chamber.
The ground foil is first laid and held by 
ground screw nuts onto the chamber. 
Each gas-bar is inserted into all the 
tube endplugs simultaneously using multiple
special brass tools and three specially 
designed jigging clamps 
(see figure~\ref{fig:gasbar-assembly-2}(a)). 
O-rings, after ultrasonic cleaning with IPA,  
are placed onto the signal caps. 
Each signal cap is inserted into each gas 
connector using a special driver shown in 
figure~\ref{fig:gasbar-assembly-2}(b).
Complete gas-bar installation on both 
multi-layers on one side of a chamber is 
shown in figure~\ref{fig:gasbar-assembly-2}(c).
\par
\begin{figure}[hbt]
    \centering
        \includegraphics[width=0.8\textwidth]{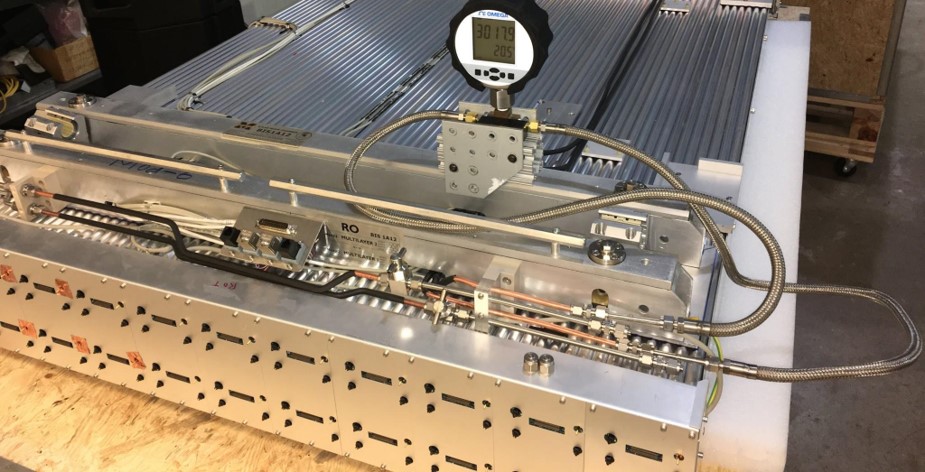}
    \caption{Completing the on-chamber gas 
        system are 6~mm stainless steel and
        copper tubing pieces,
        Swagelok valves and fittings which
        connect the gas-bars to external supply 
        lines. 
        A high precision gas pressure gauge is 
        here connected for the gas leak test, 
        and the RO Faraday cage is already 
        installed.
        }
    \label{fig:gasbar-tubing}
\end{figure}

The gas-bars are connected to an on-chamber 
gas distribution system using Swagelock 
fittings and valves connect the gas-bars as 
shown in figure~\ref{fig:gasbar-tubing}. 
A high precision gas pressure gauge is 
temporarily connected to the gas system for
chamber pressure measurements. 
The temperature of the chamber is monitored 
with 12 temperature sensors distributed on 
both sides of the chamber and readout through
the patch panel shown in the picture.
%Chamber gas leak tightness is first certified by 
%monitoring the pressure and temperature for at least %24 hours. 
%The final leak rate measurement is performed 
%after the ground-pins between the tubes and the 
%on-chamber electronics carts are installed.

\subsection{Front-end electronics installation}

\begin{figure}[!hbtp]
    \centering
    \includegraphics[width=0.8\textwidth]{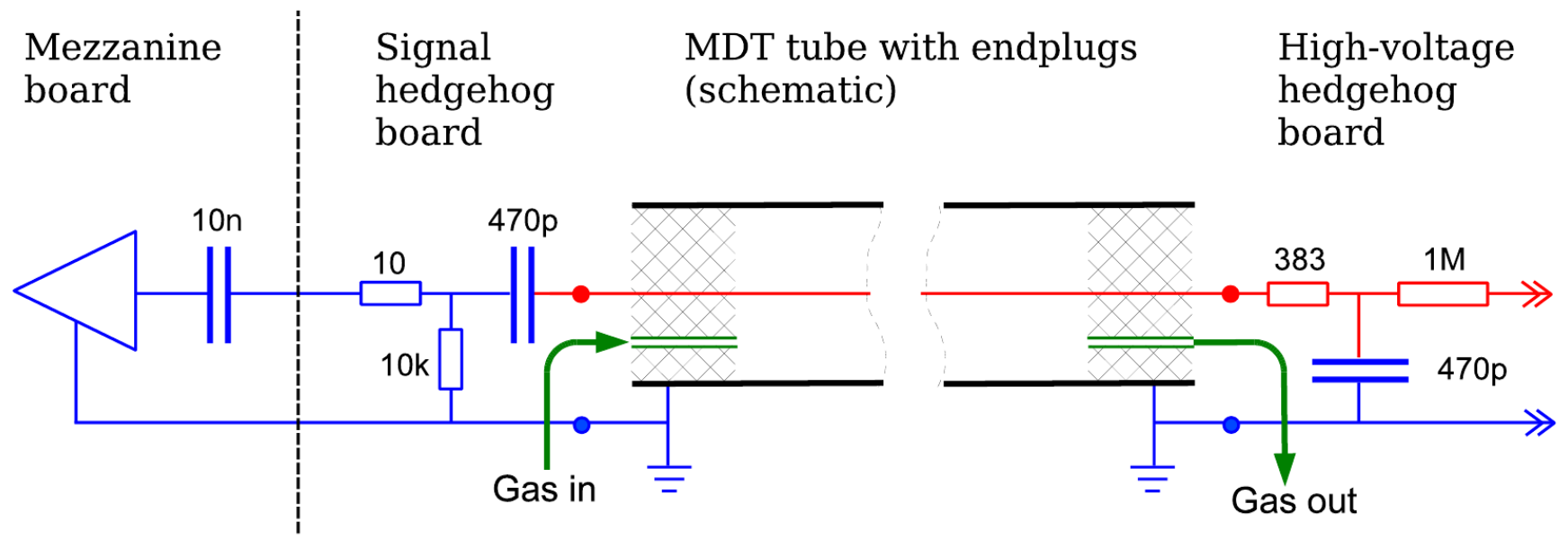}
    \caption{Electrical connections to an MDT 
    drift tube~\cite{MDTelectronics}.
    Resistors values are in ohms ($\Omega$).
    For an sMDT tube, the 383~$\Omega$ resistor 
    is replaced with a 330~$\Omega$ resistor.
    }
    \label{fig:electricalconnection}
\end{figure}

The schematic of the electrical connections 
to  an MDT drift tube is shown in 
figure~\ref{fig:electricalconnection}.
For an sMDT tube, the 383~$\Omega$ resistor 
in the HV side, shown in the figure, 
is replaced with a 330~$\Omega$ resistor.
The RO and HV hedgehog (HH) cards consist
mainly of passive components and mount directly
on the signal and grounds via sockets on the 
cards.

There are two configurations of the HH and RO 
cards. 
One type has 24 channels with configuration 
of 4$\times$6 to span the 4 layers of a ML, 
6 tubes wide.
The other has 20 channels with configuration 
of 4$\times$5, which is installed on the ends
of each ML. 
A total of 20 (16) 4$\times$6 cards plus 4 
4$\times$5 cards are installed
in the BIS1 (BIS2-6) chambers.
HV is distributed through four pairs of the 
input HV and ground cables.
The HV HH cards are chain interconnected via 
PCB jumpers.
\par
Faraday cages (FC) are installed around the
HH cards to reduce noise.  
Two HV distribution boxes are mounted on top
of the FC and distribution cables are routed 
through two holes of the FC.
The FC on the RO side has the mezzanine cable 
connection holes to allow mounting the 
mezzanine cards to read out the chamber signals.
The mezzanine cards hold the active components 
of the front-end electronics, the amplifier 
and time-to-digital converter chips.

\subsection{Chamber gas tightness measurement}

The ATLAS upper limit of the leak rate is
% \rm LR_{limit} = 
$\rm 1\times 10^{-5}
%\left[\frac{mbar ~\times ~liter}{ s} \right] 
[ mbar \, liter / s ]
\times (2 \, N_{tube}), $
where $\rm 2 \, N_{tube}$ denotes total number
of endplugs included in the measurement.
The tube gas volume is 
$\rm V(tube) \approx 250 cm^3$, 
therefore the volume of each ML is 
$\rm V_{ML} = N_{tube} \times 250~cm^3.$
Using this information, the upper limit of 
the chamber leak rate can be converted to 
a limit on the chamber pressure drop per hour 
in the measurements, which is 0.288 mbar/hour
for a ML of an sMDT chamber.
\par
Since the chamber pressure depends on gas 
temperature, the measured pressure change 
($\Delta P$) over time interval ($\Delta t$) 
must be corrected to be compared to the
specified leak rate limit.
The chamber normalized leak rate is determined 
by
$$%$$\begin{equation}
    \label{eqn:LeakR8}
    \left( \frac {\Delta P}{\Delta t}\right)_{norm} = 
    \frac{\Delta P_{corr}}{\Delta t }
$$%$\end{equation}
where $\Delta P_{corr}$ is the change in 
pressure corrected with the temperature 
variation over a time interval $\Delta t$,
calculated as:
$$ \Delta P_{corr} = 
P_{f} ~\frac{T_{ref}}{T_f} - P_{i} ~\frac{T_{ref}}{T_i} $$ 
where $P_i, ~P_f, ~T_i,~T_f$ are initial and 
final pressure and temperature, and 
$ T_{ref}=293.15^\circ$~K 
is used as reference temperature.
\par
The chamber gas output valves are connected 
to a pressure gauge (see 
figure~\ref{fig:gasbar-tubing}),
which can monitor each ML gas pressure. 
The chamber is pressurized with Helium gas, 
then a Helium sniffer is used to detect leaks
from connection points and/or individual tubes.
Most leaks can be fixed quickly by replacement
of o-rings, gas stacks, or signal caps. 
A few tubes have developed leaks which cannot 
be fixed. For these, the wire is removed, the 
end plugs sealed to isolate the tube from the 
gas system, and the signal caps modified to 
isolate the  tube from both the HV and RO
electronics.
\par
\begin{figure}[hbt]
    \centering
    \includegraphics[width=0.78\textwidth]{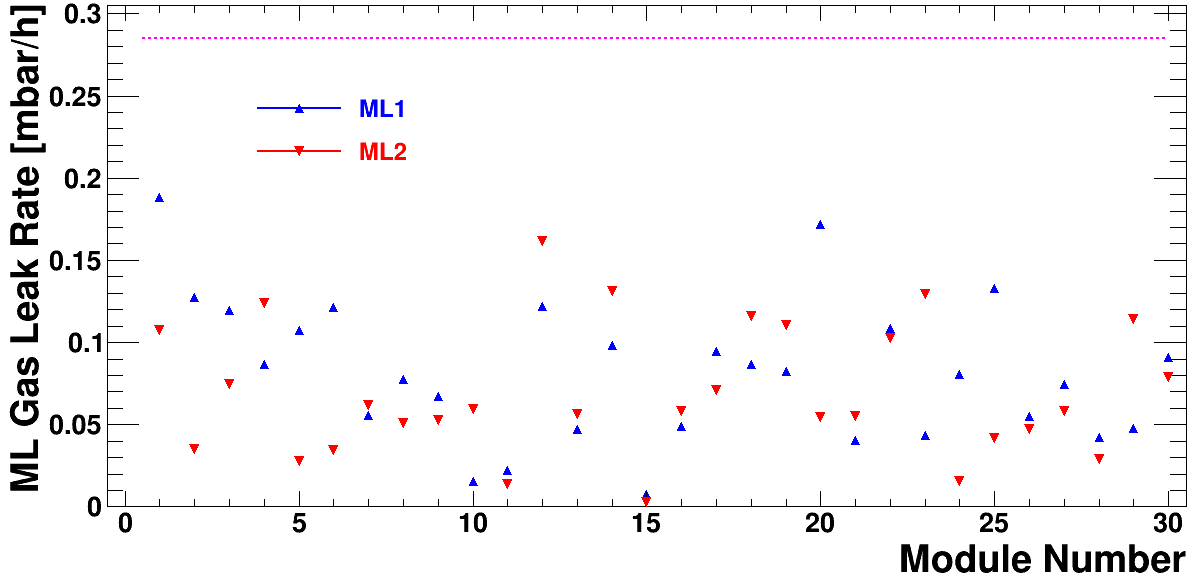}
    \caption{Multi-layer gas leak rate of the first
        30 sMDT chambers built at UM.
        The dashed line represents the ATLAS 
        maximum acceptable gas leak rate.}
    \label{fig:ChamberLeakRate}
\end{figure}

The final leak rate measurement is performed 
with all HH cards and the Faraday cages 
installed.
The resistance of all 12 temperature sensors 
is measured in Ohms, converted into degrees 
Celsius and averaged for each ML to correct 
the initial and final chamber ML pressures
readout during the leak test, which check 
the pressure drop as a function of time.
The measured ML leak rates of the first 30 
sMDT chambers built at UM are shown in
figure~\ref{fig:ChamberLeakRate}. 
The pink dashed-line in the plot indicates the 
upper limit, and blue (red) triangles are the 
measured leak rates for ML1 (ML2) of produced 
chambers at UM, all well below the allowed 
upper limit. 
%Once the leak test is completed, the Helium gas is purged, and the chamber is flushed with ten volumes of drift gas mixture, $\rm Ar:CO_2 ~(93:7)$, in preparation for the electronics testing. 

\subsection{Ground cable connections}

\begin{figure} [!hbt]
    \centering
    \includegraphics[width=0.98\textwidth,height=\linewidth,keepaspectratio]{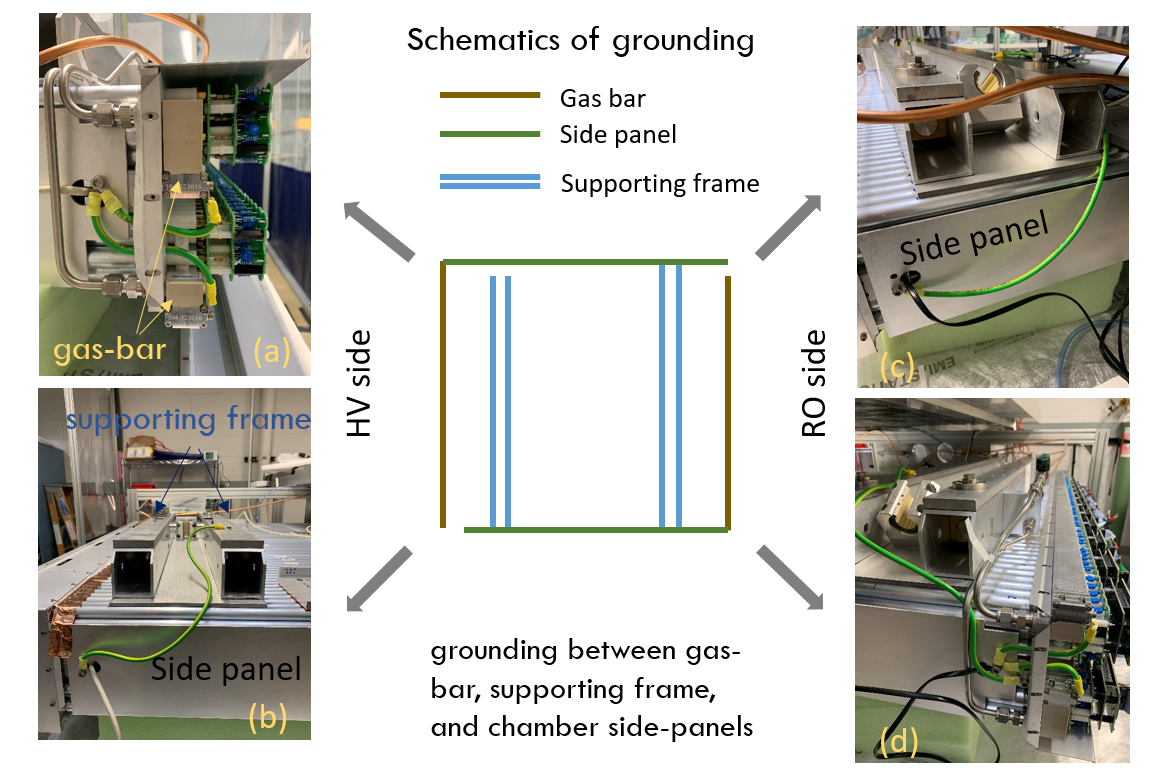}
      \caption{Grounding cable connections 
      schematics.
      Pictures show detailed connections to the 
      gas-bar (a), side-panel (b), support 
      structure (c), and between the gar-bar and 
      the support structure. 
    }
    \label{fig:groundwires}
\end{figure}

The chamber grounding follows a "star"
design that avoids ground-loops.
The schematic of the ground connectivity
is shown in figure~\ref{fig:groundwires},
where the pictures show how the 
grounding cables (in green color)
connect the gar-bars, 
chamber side-panels, and the chamber 
support structures.
%To avoid a grounding loop the grounding wires are 
%installed on a chamber without a connection 
%between the support structure and the side panel %on the HV side (near the gas inlet). 

\section{Electronics and Cosmic Ray Test}
\label{sec:cosmic}
The final QA/QC tests include measurements
of the noise level and efficiency of each tube,
as well as the chamber tracking efficiency and
resolution.
These tests are carried out in the cosmic ray 
test station which has been 
described in Section~\ref{sec:cosmicTestStation}.
\begin{figure}[!hbtp]
    \centering
    \subfloat[]{
        \includegraphics[width=0.44\textwidth]{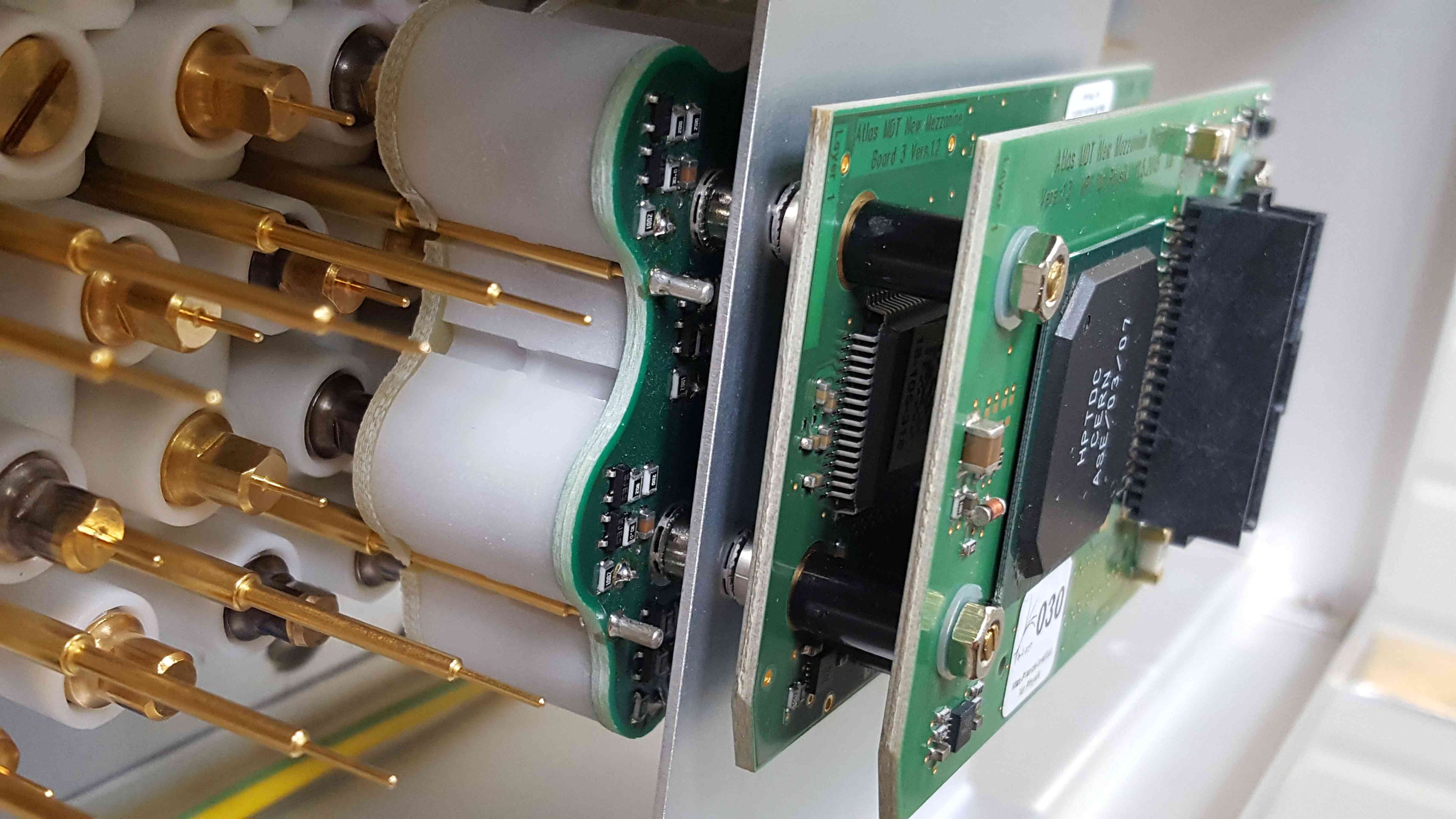}
    }
    \subfloat[]{
        \includegraphics[width=0.55\textwidth]{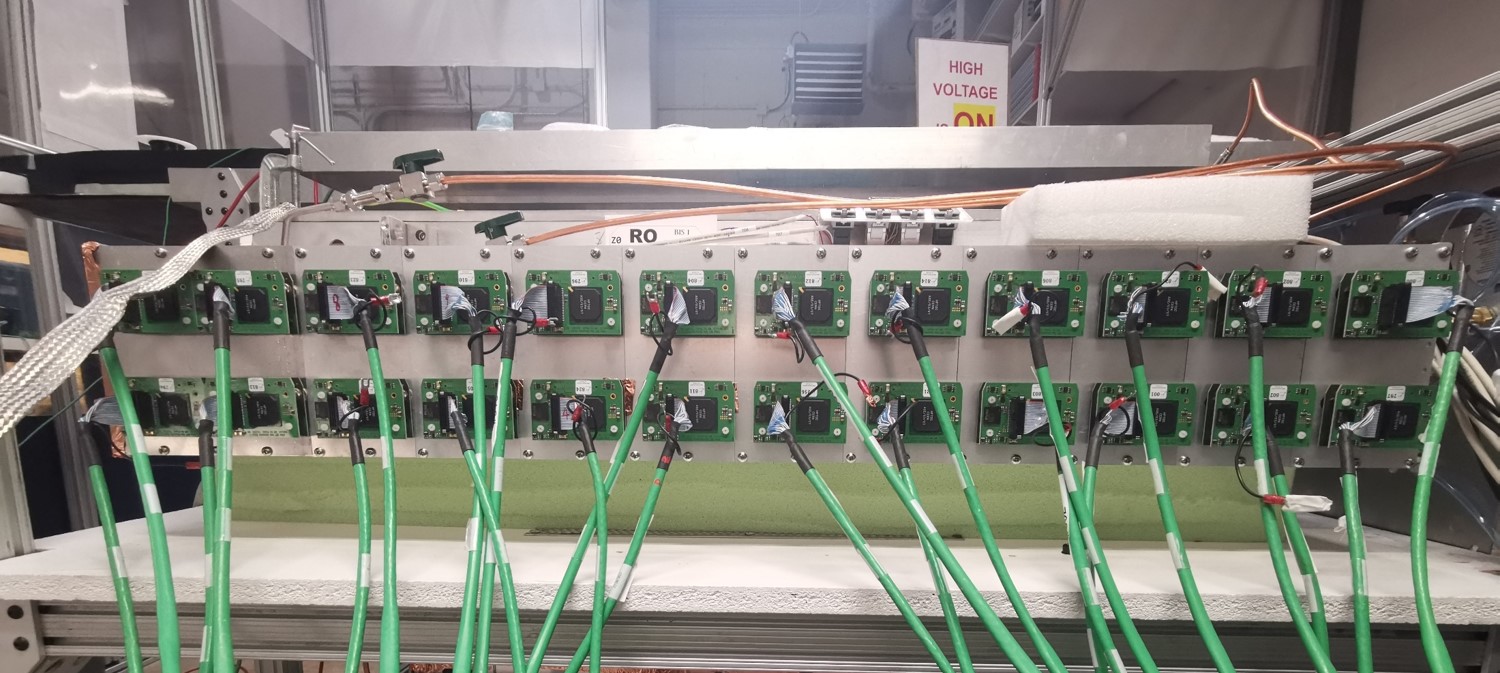}
    }
    \caption{(a) A 24 channel RO HH and mezzanine 
        card mounted on a sMDT chamber;
        (b) 24 front-end electronics cards 
            on a BIS1 chamber RO side.}
    \label{fig:mezz}
\end{figure}

Each chamber runs for a minimum of 24~hours at 
the operating HV (+2730~V) to ensure that the
current drawn is  at a level below 2~nA times
number of tubes.
Tubes which draw excessive current are 
individually treated with negative HV 
(up to -3~kV) for about one hour to reduce
the current below the 2~nA limit. 
Finally, the readout mezzanine cards are 
installed on top of the RO HH cards.
One mezzanine card contains a board with three 
ASD (Amplifier/Shaper/Discriminator) ASIC chips,
each with 8~channels, and another board with one
24 channels high-performance time-to-digital
converter (HPTDC) chip on top of the ASD board.  
Figure~\ref{fig:mezz}(a) shows a mezzanine card 
installed on a chamber.
Two Chamber Service Modules (CSMs) are used to 
multiplex the readout from all the HPTDCs of a 
chamber (each CSM can multiplex up to 18 HPTDCs).
The DAQ system with a scintillator trigger and 
the online monitoring system are connected to 
the test chamber for measurements.
An offline analysis program 
(detailed description of the algorithms
in~\cite{MichiganSMDTResolutionJINST})
was developed to measure the performance of each
tube and ultimately the efficiency and the 
resolution.

\subsection{Noise rate measurements}

Each tube noise rate is measured in a 
series of data runs at different readout 
thresholds with both HV-off and HV-on 
(at the nominal working voltage of +2730 V).
A 10~KHz pulse-generator is used as trigger 
to read out noise hits. 
The noise rate can be calculated based on the 
recorded number of hits and the number of 
triggers during the data taking period
(typically $\Delta t$ = 10 minutes) as:
$$ Noise \; rate  = \frac{ N_{hits} }{ N_{trigger} }  \times 
\; time_{window} \; , $$
where $ {time}_{window}  (1.55 \; \mu$s) 
is the HPTDC readout time window; 
$ N_{hit} $ is the number of hits in each channel 
recorded in the test time $\Delta t$, and  
$ N_{trigger} = 10 {\rm ~KHz}\times \Delta t $.
\begin{figure}[htb]
    \centering
    \subfloat[]{
    \includegraphics[width=0.51\textwidth]{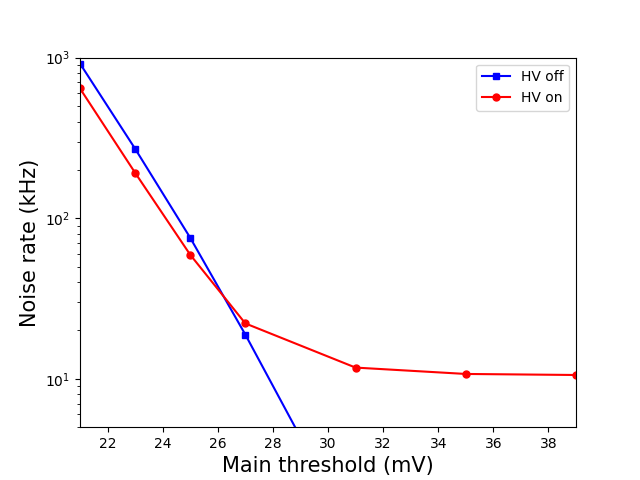}
     }
    \subfloat[]{
    \includegraphics[width=0.47\textwidth]{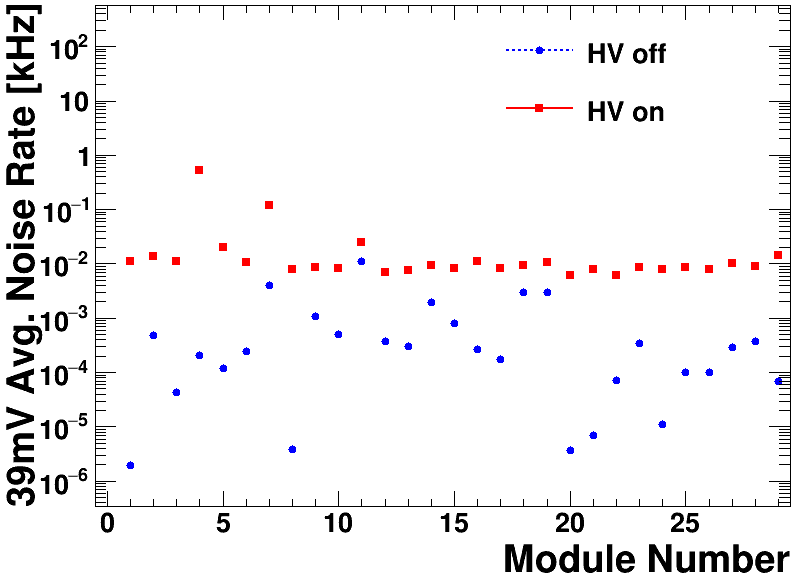}
     }
    \caption{(a) Noise rate vs. threshold for a 
        representative sMDT chamber. 
    (b) Chamber average noise rate measured at 
        39~mV threshold for the first 30 modules.
        }
    \label{fig:NoiseMax_39mV}
\end{figure}
\par
The measured noise rate vs. ASD threshold for 
a representative sMDT chamber built at UM is 
shown in figure~\ref{fig:NoiseMax_39mV}(a). 
When the readout threshold is larger than 
$\sim$30~mV, the chamber hit rate is dominated
by the cosmic rays, as seen from the red curve 
measured with HV-on.
The specification of the maximum tube noise rate
is 500 Hz at the readout threshold of 39~mV.
Figure~\ref{fig:NoiseMax_39mV}(b) shows
the average noise rate measured at 39~mV 
threshold for the first 30 sMDT built at UM,
all well below 500 Hz, except Module-4 with 
noise rate close to the specification in the 
HV-on measurement. This chamber will be closely 
evaluated again during the final commissioning 
phase at CERN. 

\subsection{Tube spectra from cosmic ray data}

\begin{figure}[hbt]
    \centering
    \subfloat[]{
    \includegraphics[width=0.49\textwidth]{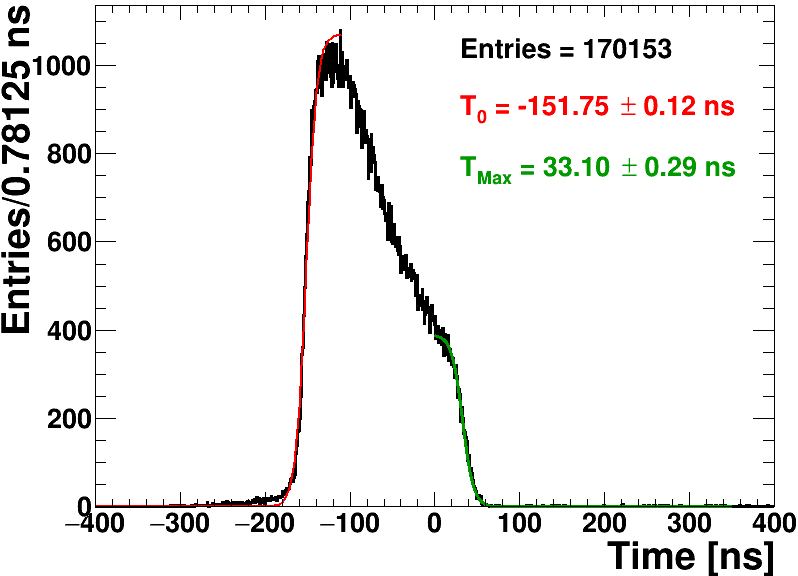}
    }
    \subfloat[]{
    \includegraphics[width=0.49\textwidth]{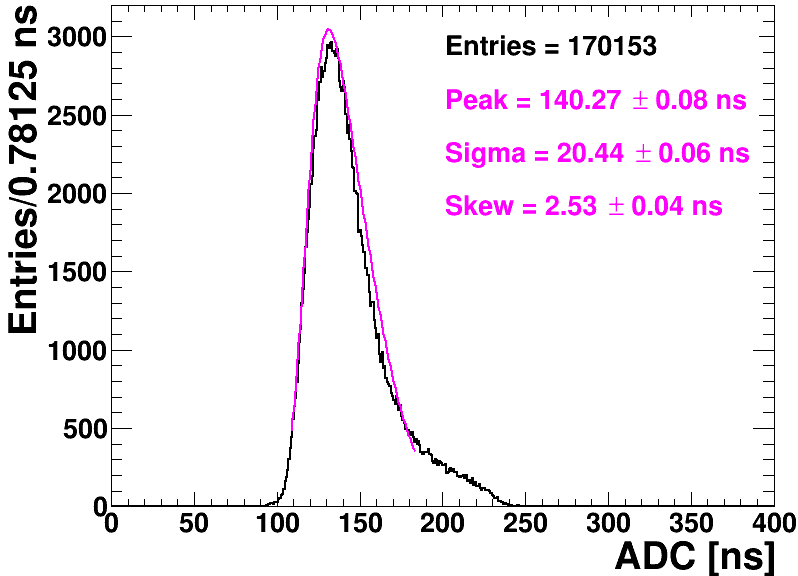}
    }
    \caption{Distributions of cosmic ray
        hits recorded by a typical tube: 
        (a) time, (b) charge.}
    \label{fig:ADC-TDCspectra}
\end{figure}

Each tube's response to charged particles
is tested with cosmic rays.
Figure~\ref{fig:ADC-TDCspectra} shows typical 
(a) time (TDC) and (b) charge (ADC) spectra 
for the hits of a tube on a representative 
sMDT chamber together with their fits (described 
in~\cite{MichiganSMDTResolutionJINST}).
Both the leading and trailing edge of the TDC 
distribution are fit with a Fermi-Dirac function
in order to get the time offset ($T_0$) and
the maximum time $T_{Max}$.
The ADC distribution is fit with a skew 
normal function.

The measured hit drift time, calculated as 
the difference between the hit time and the
$T_0$, is then corrected with the time-slew 
correction using the charge information.
The time-slew effects are due to the jitter 
introduced by different times of crossing the 
ASD threshold by different amplitude signals.

\subsection{Tube and chamber efficiencies}

\begin{figure}[hbt]
    \centering
    \subfloat[]{
        \includegraphics[width=0.46\textwidth]{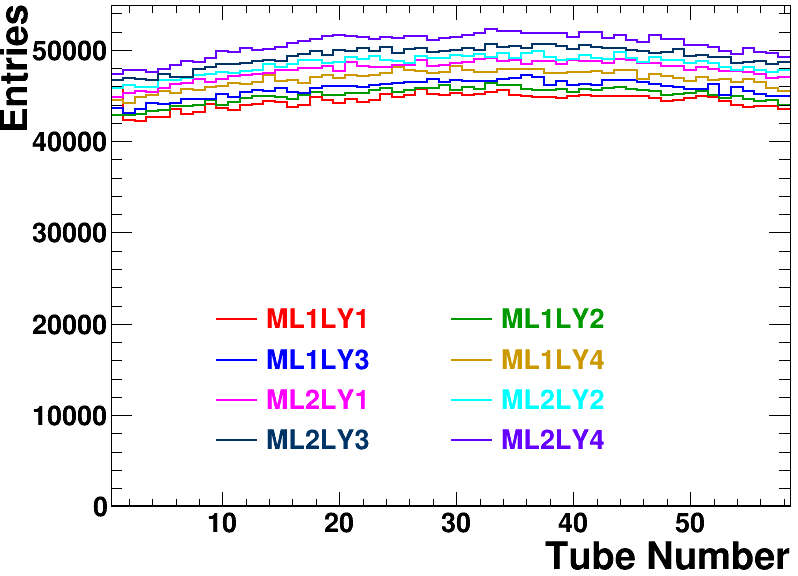}
    } 
    \subfloat[]{
        \includegraphics[width=0.51\textwidth]{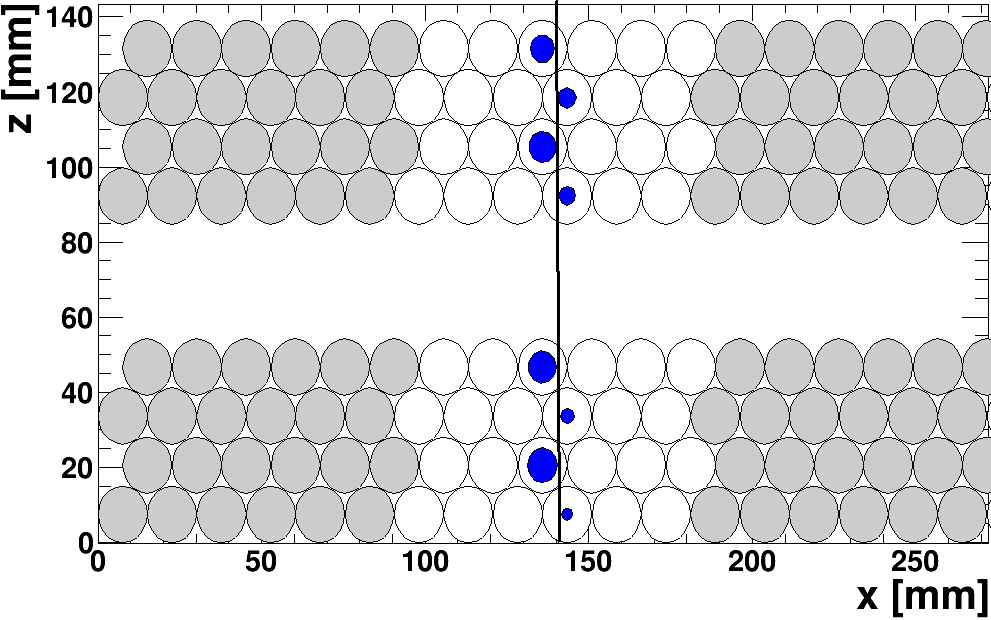}
    }
    \caption{
    (a) Cosmic ray run hit map for the 8 tube
    layers as a function of the tube number 
    for a representative sMDT chamber;
    (b) example of a track reconstructed 
    in a sMDT chamber by a fit tangent to the
    drift circles (blue circles).
    Grey (white) circles are drift tubes read out 
    by a different (the same) mezzanine card.
    }
    \label{fig:ChamberHitMapTrack}
\end{figure}

Figure~\ref{fig:ChamberHitMapTrack}(a) shows
an example of the chamber response uniformity 
based on the number of hits per tube in the 
cosmic ray run.
\par
The definitions of efficiency rely on having
successfully reconstructed a track.
A muon track is built starting from the sMDT 
hits for each triggered event by fitting the 
hits drift distances obtained from the 
(corrected) drift times via an $R(t)$
function. 
This time-to-space relationship is obtained 
with an auto-calibration program that converts 
the drift time to the drift distance through a procedure of minimizing the tracking residuals 
of all the collected muon tracks in a test run.
An example of a comic ray muon track reconstructed
in an sMDT chamber built at UM is hown in
figure~\ref{fig:ChamberHitMapTrack}(b).

\begin{figure}[hbt]
    \centering
    \subfloat[]{
    \includegraphics[width=0.49\textwidth]{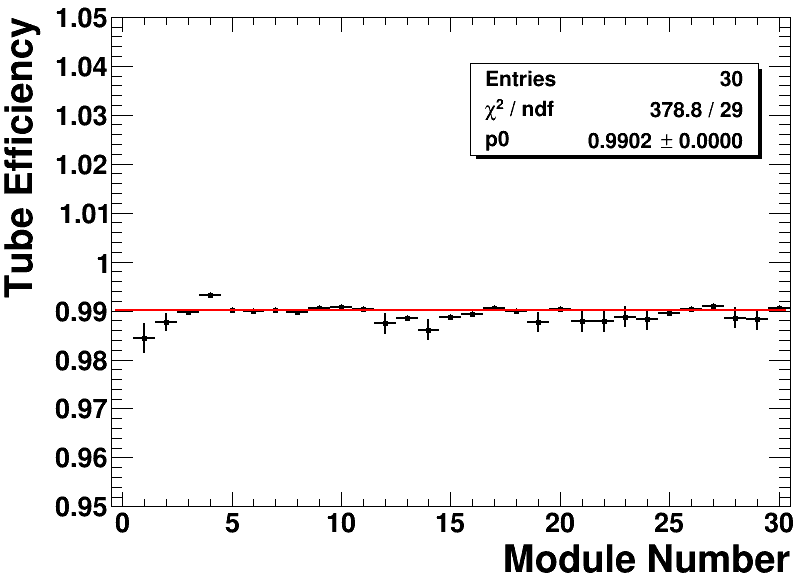}
    }
    \subfloat[]{
    \includegraphics[width=0.49\textwidth]{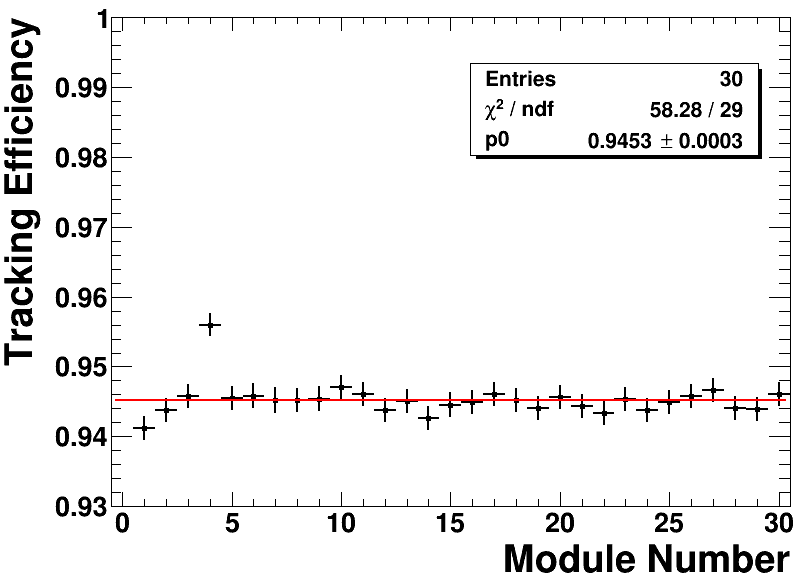}
    }
    \caption{(a) Tube and 
             (b) tracking efficiency for the first 
                 30 sMDT chambers built at UM.}
    \label{fig:Efficiency}
\end{figure}

The tube efficiency is defined as the 
ratio between the number of hits recorded on a
track passing through that tube and the number 
of reconstructed tracks passing through the
gas-volume of that tube.
Figure~\ref{fig:Efficiency}(a) shows this 
single tube detection efficiency, which is on 
average $\rm 99.02 \pm 0.01 $\% for the first 
30 chambers built at UM.
\par
The tube layer efficiency is defined as the 
number of hits in that layer belonging
to a track divided by the number of tracks
passing through that layer. 
This latter efficiency includes the tube
efficiency as well as the inefficiency of 
the tube walls and the space between tubes
(dead regions).
Figure~\ref{fig:Efficiency}(b) shows the 
tube layer efficiency averaged over a chamber 
(called tracking efficiency).
The average tracking efficiency for each 
tube layer is  
$94.52 \pm 0.03 $~\%.
\par
%4480 BIS1 (70x8LYx8sMDT) + 102-8 BIS2-6(58x8LYx22sMDT)
Out of the 14688 tubes of the first 30 UM sMDT 
chambers, 13 were lost due to wire slippage
(insufficient crimping) or irreparable gas 
leaks (cracked endplug or improper endplug 
swaging).

\subsection{Tracking resolution}

The ATLAS muon spectrometer measures the momenta 
of muons by reconstructing the curvatures of their
trajectories in the magnetic field, and thus
the spatial tracking resolution will in turn 
determine the resolution of the momentum 
measurement.
\par
\begin{figure}[hbt]
    \centering
    \subfloat[]{
    \includegraphics[width=0.49\textwidth]{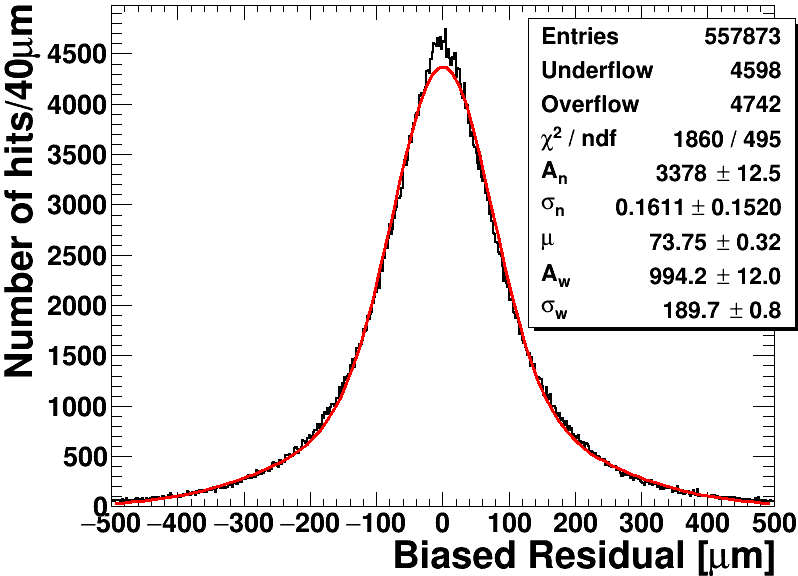}
    }
    \subfloat[]{
    \includegraphics[width=0.49\textwidth]{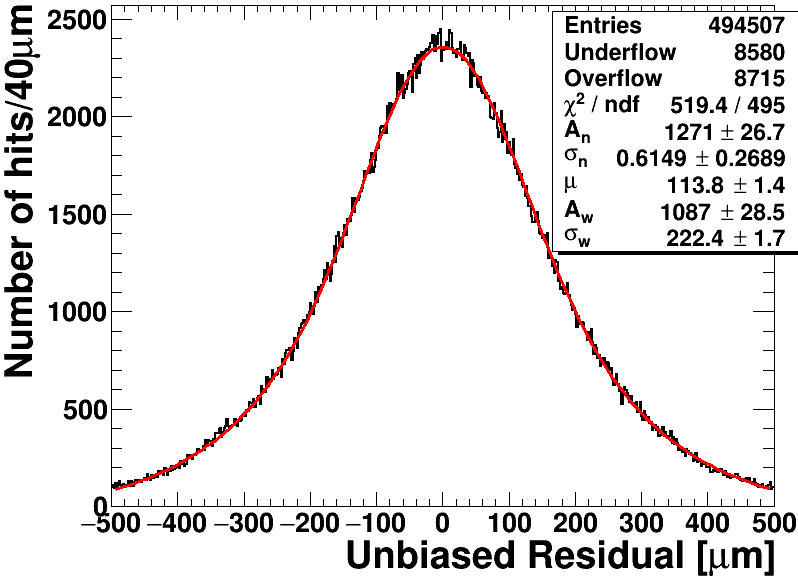}
    }
    \caption{(a) Biased and (b) unbiased residual   
        distribution for a representative sMDT
        chamber built at UM with a double Gaussian
        function fit.}
    \label{fig:residues}
\end{figure}

The single tube resolution is determined by fitting 
the reconstructed muon tracks and evaluating the
residual distributions.   The tracking residual is
the difference 
$ r_{\textrm{drift}} - r_{\textrm{fit}} $,
where $ r_{\textrm{drift}}$ is the measured drift 
radius and $ r_{\textrm{fit}}$ is the radius 
obtained by the track fit. 
Two types of residuals are introduced, biased 
and unbiased.
The biased one is the residual when the track
is fit with all hits.
The unbiased one is the tracking residual of a 
hit from a track reconstructed by removing that
hit from the fit.
Figure~\ref{fig:residues} shows the biased and unbiased residual distributions. 

A residual distributions is fitted with a double 
Gaussian function with common mean
(see~\cite{MDTResolution} for details) 
and the single wire resolution is defined as \cite{Carnegie}:
\begin{equation}
\sigma_{wire} = \sqrt{ \, \sigma_{B}^{} \cdot \sigma_{U}^{}\, }
\end{equation}
where $\sigma_{B}^{} \; (\sigma_{U}^{}) $ is the 
biased (unbiased) residual width calculated as 
amplitude weighted average of the two Gaussian
components, narrow ($n$) and wide ($w$), 
of the fit:
\begin{equation}
\sigma_\alpha = \frac{ A_{n} \cdot \sigma_{n} + 
                   A_{w} \cdot \sigma_{w} }{ A_{n} + A_{w} }, \hspace{1cm} (\alpha=B, \ U).
\end{equation}
%ChamberResolution
\begin{figure}[hbt]
    \centering
    \subfloat[]{
    \includegraphics[width=0.49\textwidth]{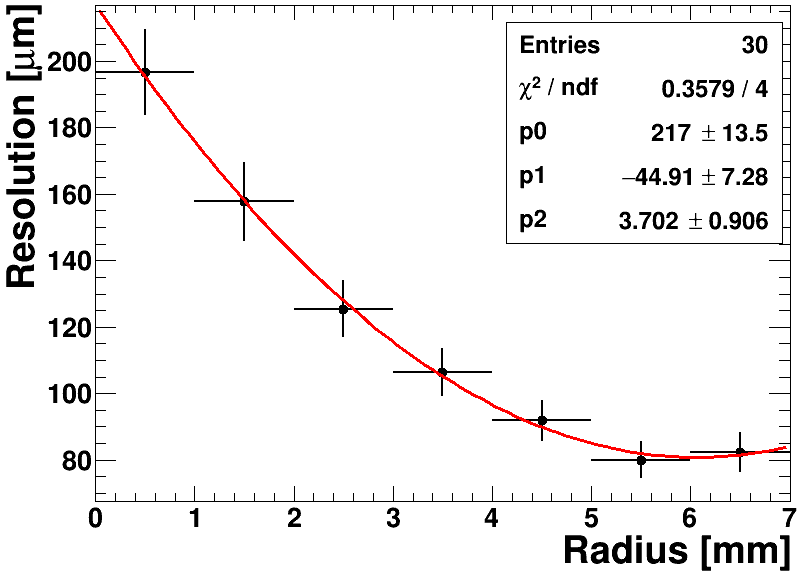}
    }
    \subfloat[]{
    \includegraphics[width=0.49\textwidth]{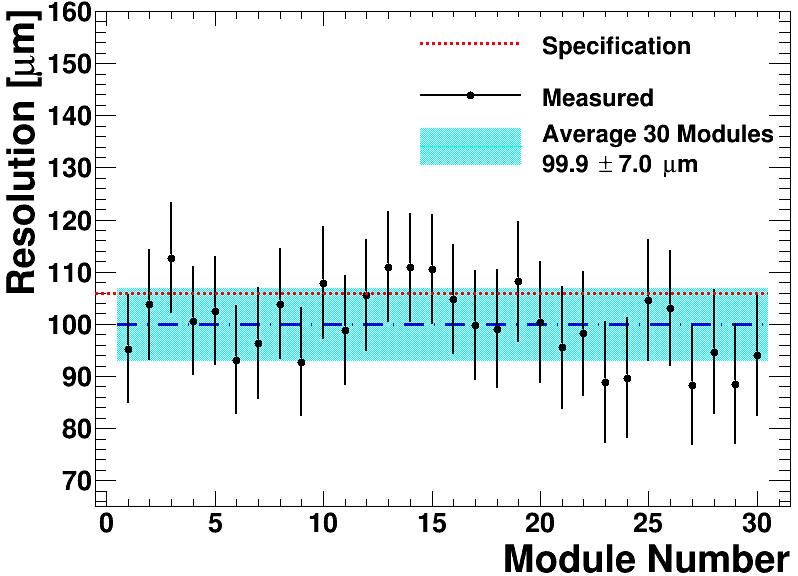}
    }
    \caption{
    (a) Average resolution vs. radius of the
        30 sMDT chamber built at UM.
        The errors are the standard
        deviations of the 30 data-point
        at that drift radius;
    (b) chamber resolution for the first 30 
        sMDT chamber built at UM.
    }
    \label{fig:ChamberResolution}
\end{figure}

The multiple Coulomb scattering effect on low 
energy cosmic ray muons are corrected when 
computing the resolution~\cite{MichiganSMDTResolutionJINST}.
The single wire resolution vs. the tube radius 
for a representative sMDT chamber built at UM
is shown in figure~\ref{fig:ChamberResolution}(a).
The measured tracking resolution (integrated 
over all radii) for the first 30 chambers made 
at UM shown in 
figure~\ref{fig:ChamberResolution}(b),
is $ 100 \pm 7$~$\mu$m to be compared 
with the expected value of 
106~$\mu$m~\cite{smdtDesign}.
The resolution of the constructed chambers
is better than the specification because the
expected value was estimated using the first
generation of the electronics (in particular
the ASD-1), while the sMDT built at UM are 
tested with mezzanine cards instrumented 
with the final ASD-2 which has a higher gain.

All chambers constructed at UM so far meet the performance specifications both for 
tracking resolution and efficiency.

\section{Database}
\label{sec:database}
All the information about each chamber construction 
precision measurements, gas tightness, and 
electronics tests as well as the overall 
performance are assembled into a local database,
ready to be checked and/or displayed.
The data from the various measurements and tests 
stations is usually saved in text files, either
directly or after an initial decoding/analysis 
on the output data from a test station.
This format of the local data storage allows converting to different databases 
(MS Access, MySQL, Oracle, JSON, ...) for the task.
The most natural choice is a relational database, 
since all information to be saved is relative to 
many single units (as tubes, sensors, plates, 
tube-layers, multi-layers, ...), each one with a 
list of properties, test parameters and results: 
all data is correlated.
\par
With the goal to store, quickly retrieve, and 
eventually show the data, a ROOT~\cite{root} 
based approach is used.
C++ classes read each chamber (and corresponding
tubes) data files and organize them into a
(multiple) set of measurements with conditions
and relative data into a {\it TTree} of the 
ROOT file.
The main blocks of this {\it TTree} are described 
below.
\begin{description}
   \item[Construction:] Includes the IDs of the 
   tubes, their locations on the chamber and the 
   measurements of the tube heights relative to
   the granite table.
   From this data the average height of a 
   tube layer is calculated and then each layer
   and spacer frame Y-pitch are determined.
   Also, the positions of all the platforms (AP, 
   B-field, and CCC) glued on the chamber are 
   recorded.
   \item[In-plane alignment:] Stores the RASNIK 
   system data from the final measurement on the 
   granite table (the base-line measurements) and 
   at different inclination angles on the rotation
   cart.
   \item [Gas leak rate:] Saves the gas pressure
   for each chamber ML together with the
   temperature at different times to determine
   the gas leak rate.
   \item [Noise level:] Retains the noise rate
   for each tube for HV-off and HV-on runs at
   8 different ASD thresholds.
   \item [Electronics performances:] Collects
   each tube efficiency together with its cosmic 
   ray test TDC and ADC spectra, as well as
   the tracking efficiency, the chamber biased
   and unbiased hit residual distributions,
   the single wire resolution, and the 
   resolution vs drift radius.
\end{description}
In all these ROOT-based DB {\it TTrees}, the 
tube barcode ID is the primary key for 
retrieving information from the measurements
on a tube, either as a single unit 
(wire tension, dark current, and so on, see~\cite{UMtubePaper})
or part of an assembled chamber (tube heights 
over the granite table surface, efficiency, 
resolution, and so on.)
%\par
While filling the {\it TTree}, a set of standard 
plots for each test is automatically saved, ready 
to show trends and/or discrepancies between a
chamber and all the others already tested or
reference values.
A dedicated set of files with common formats, 
which are followed by both Michigan and the MPI
production sites, is prepared and uploaded to 
the CERN website\footnote{https://atlas-mdt-phase2-qc.web.cern.ch/}
to share the results by both sites on the sMDT 
chamber production quality control. 
This common database on sMDT production will 
be used for future sMDT commissioning and 
operations in the ATLAS experiment. 

\section{Conclusion}
\label{sec:conclusion}
%\section{Conclusion}
%\arabic{nChambers}

With the elaborate infrastructure built at UM,
30 sMDT chambers have been constructed meeting
the stringent precision and quality requirements.
These chambers are all thoroughly tested with 
cosmic rays and achieve an single tube efficiency
above 99\%, and tracking resolution of
$\rm 100 \pm 7 \; \mu m $.
\par
There are three major steps in the chamber 
construction and certification (base chamber 
gluing, services installation, and final 
cosmic ray testing), each requiring 10 working 
days on average.
The three steps are carried out in parallel in 
order to meet the production schedule.
\par
Of the 30 chambers constructed at UM, the first 
batch of 24 chambers has already been delivered
to CERN.
20 more chambers are still to be constructed 
at UM and will be delivered to CERN by the 
end of 2023.
The final commissioning of all 50 Michigan 
chambers prior to installation on the ATLAS 
cavern is planned to begin in 2024 at CERN.

\acknowledgments

We would like to thank  
Hubert Kroha and Oliver Kortner (MPI), 
Rinat Fakhrutdinov (IHEP, Protvino), 
Reinhard Schwienhorst (MSU), 
Tristan Du Pree (NIKHEF), 
Pierre-Fran\c{c}ois Giraud (Saclay), 
and their team members for their close 
collaborations and valuable technical 
discussions.
We also thank the UM ATLAS faculty, 
Jianming Qian and Shawn McKee for their strong 
support, and our undergraduate students, 
Tyler Coates, Kathryn Ream, Dylan Ponman, 
Saarthak Johri, and Yuxin Wan, for their 
great technical assistance.
\par
This work was supported in part by the U.S. 
Department of Energy grant DE-SC0012704 
and in part from the U.S. National Science 
Foundation PHY-1948993. 

\bibliographystyle{JHEP}
\bibliography{bibliography}

\end{document}